\documentclass[a4paper,fleqn]{article}
\usepackage[dvips]{graphics}
\usepackage{amsmath}   
\usepackage{a4,amssymb,epsfig}

\numberwithin{equation}{section}

\newcommand{\la}{\left\langle}
\newcommand{\ra}{\right\rangle}

\newcommand{\lla}{\langle\!\langle}
\newcommand{\rra}{\rangle\!\rangle}
\newcommand{\p}{\partial}

\begin{document}
\title{Storage option an Analytic approach}
\date{}
\author{}
\maketitle

\begin{center}

Dmitry Lesnik \\
\today

\end{center}


\begin{abstract}
The mathematical problem of the static storage optimisation is formulated and solved by means of a variational analysis. The solution obtained in implicit form is shedding light on the most important features of the optimal exercise strategy. We show how the solution depends on different constraint types including carry cost and cycling constraint. We investigate the relation between intrinsic and stochastic solutions. In particular we give another proof that the stochastic problem has a ``bang-bang'' optimal exercise strategy. We also show why the optimal stochastic exercise decision is always close to the intrinsic one.

In the second half we develop a perturbation analysis to solve the stochastic optimisation problem. The obtained approximate solution allows us to estimate the time value of the storage option. In particular we find an answer to rather academic question of asymptotic time value for the mean reversion parameter approaching zero or infinity. We also investigate the differences between swing and storage problems. The analytical results are compared with numerical valuations and found to be in a good agreement.
\end{abstract}


\section{Introduction}

The problem of storage optimisation is driven by the necessity of the storage owners to optimise their expenses and maximise a potential profit which can be gained by operating the storage. There are plenty of real world storage examples: Gas storage, Oil storage, Hydro power plant, Coal stock, etc. By following some clever strategy -- buying the underlying commodity cheep, storing it, and selling as the prices go up -- the storage owner can make a profit. Thus, there is a need for the exercise strategy optimisation -- such an exercise rule, which allows to maximise the profit and minimise the risks.

One of the most popular ways of the numerical storage optimisation is the ``dynamic programming'' algorithm. It allows to treat both -- intrinsic (when the prices are supposed to be frozen) as well as stochastic (when the prices are supposed to follow some stochastic process) problems. Usually the numerical solution provides the answers to the two most important questions: what is the optimal exercise decision now given the current state (prices and volume level of the storage), and what is the expected profit, provided we follow the optimal exercise strategy. The next level of sophistication is to provide the hedge strategy -- a portfolio of derivatives (futures, options or any other financial instruments) which would minimise the financial risks. Of course, financial risk is only a feature of the stochastic problem, as the solution of the intrinsic problem is deterministic.

In this paper we to develop an analytical approach to the storage optimisation problem. In the Sec.~\ref{sec:Deterministic_problem} we consider an intrinsic problem which deals with predefined deterministic price curve. Section~\ref{sec:other_constraint_types} considers different special cases of constraints. Sec.~\ref{sec:stochastic_problem} is devoted to the stochastic problem, where we make use of perturbation theory to solve the stochastic problem, and derive an estimate of the stochastic time value. In the sections~\ref{sec:example_calculation_storage} and~\ref{sec:swing_option} we make an example calculation of the time value of a simple storage and swing options and compare the results with numerical evaluation. We discuss the results in Sec~\ref{sec:discussion}.

\section{Problem formulation}

The storage problem can be formulated as follows. The storage option holder is given a right to store some amount of underlying (let it be for simplicity gas) in a (virtual) storage facility. At every time moment the option holder may ``do nothing'', inject or release the gas from the storage. Every time the gas is injected into the storage it must be bought on the market. Likewise every time the gas is released from the storage, it is sold on the market. Since the market price of gas changes with time, this may lead to a non-trivial cash-flow. The injection and release process must satisfy some operational constraints (for instance maximum injection/release rates, storage capacity, etc.), specified as boundary conditions. Every exercise profile (trajectory in the time-volume space) yields a different profit. In this sense the profit becomes a functional on the exercise trajectory. The aim of the storage option holder is to maximise the profit by choosing an optimal injection/release strategy. The problem can thus be formulated in terms of variational analysis -- an optimal trajectory is the one delivering maximum of the profit functional.

If the market prices are known in advance and never change, such a problem is deterministic. The possible profit is bound from above and from below, and thus there exists a trajectory such that no other trajectory yields higher profit. The maximal profit of the static problem is called ``intrinsic value''. In the current section we investigate the deterministic problem by means of variational analysis.

Let the storage time span be $t\in[0, T_e]$. Let $F(t)$ be the market price of gas by the time $t$. The curve $F(t)$ is called the forward price curve, since it is observed on the market prior the time $t=0$ and  contains the information about the prices of gas with delivery in the future. By the definition of the static problem the forward curve never changes, and hence it does not depend on the observation time.

Let $q(t)$ be the amount of gas in the storage by the time $t$. The curve $q(t)$ defines the exercise trajectory. The initial and terminal conditions are
\begin{align}
     &q(0)= Q_{start}\,;  \\
     &q(T_e)= Q_{end}\,.
\end{align}

The cash flow (per unit time) resulting from trading the gas according to the exercise strategy $q(t)$ is given by
\begin{eqnarray}
    \label{eq:1.1}
    - \dot q(t)\,F(t)\,.
\end{eqnarray}
An additional cash flow results from taking into account the injection/release costs (operating costs). Let us designate $\gamma(\dot q(t))$ the operating costs per unit time. The terminal profit is given by the cumulative cash flow. We thus introduce the target functional
\begin{eqnarray}
    \label{eq:unperturbed_action}
    S_0[q(t)] = -\int_0^{T_e} \Big[ \dot q(t)\,F(t) + \gamma(\dot q(t)) \Big]\,dt\,.
\end{eqnarray}
The value of this functional on the trajectory $q(t)$ gives the storage value conditional on that trajectory. We have used here subscript $0$ to indicate the unmodified action integral. In the next section we will introduce a modified action integral $S[q(t)]$, which includes additional terms intended to enforce the operational constraints. Running ahead we notice that on any trajectory $q(t)$ allowed by constraints the values of the modified and unmodified functionals coincide.

Using ``physical'' terminology we introduce the (unmodified) Lagrangian $L_0$ corresponding to the target functional $S_0$ as
\begin{align}
    \label{eq:unperturbed_lagrangian}
    L_0 = - \dot q(t)\,F(t) - \gamma(\dot q(t))\,.
\end{align}

As mentioned above, the target functional is bound from above and from below, and hence there must exist such trajectories on which the functional achieves its maximum and minimum\footnote{We do not conduct a thorough analysis of existence and uniqueness of the solution of the variational problem. However we point out that the class of functions $q(t)$ on which the target functional is defined and could reach maximum should be rather broad. In particular it must include all continuous locally integrable functions. Other functions under the functional integral are allowed to be discontinuous. We will also use the concept of convergence of functions, which we will always understand as a weak convergence, i.e. $\psi_k\to\psi$ if $S[\psi_k]\to S[\psi]$.}. We make use of variational analysis to search the extremal trajectory, i.e. the trajectory on which the first variation of the target functional vanishes. Once the extremal trajectory is found, one has to make sure that it delivers the maximum of the functional. It could be done by evaluating the second variation and checking its sign.

The operational constraints may differ for different storage option types. Below we consider some typical constraints, which can be classified as \emph{local}. Local property implies that a constraint at time $t$ can be expressed in terms of state variables and their derivatives $q(t),\dot q(t), F(t),\dot F(t), ... $ at time $t$.

We consider the following two operational constraints:
\begin{enumerate}
    \item The volume $q(t)$ is allowed to be within the interval
    \begin{eqnarray}
        q(t) \in [Q_{min}(t),Q_{max}(t)]\,,
    \end{eqnarray}
    where the boundaries $Q_{min}(t),Q_{max}(t)$ are time dependent.
    \item The injection/release rate $\dot q(t)$ is bounded to
    \begin{eqnarray}
        \dot q(t) \in [r_{min}, r_{max}]\,.
    \end{eqnarray}
    Generally the maximal injection/release rates may depend on the time and volume: $r = r(t,q(t))$. Below we only consider the constant injection/release rates. The special case of volume dependent rates will be considered in the Sec.~\ref{sec:volume_dependend_rates}.
\end{enumerate}

One of the most common examples of \emph{nonlocal} constraints is the so called \emph{cycle} constraint. It can be formulated as follows. One introduces an \emph{intake} cycle variable as
\begin{align}
    \label{eq:cycle_variable}
    c(T) = \int_0^{T} \dot q(t)\, \theta( \dot q(t) )\,dt\,;\qquad\text{where} \quad
    \theta(x) = \left\{
        \begin{array}{ll}
            1\,,\quad & x\geq 0 \\
            0\,,      & x < 0
        \end{array}
    \right.
\end{align}
which has a meaning of total injected volume by the time $T$. The cycle constraint requires that the terminal value $c(T_e)$ does not exceed certain threshold
\begin{align}
    \label{eq:cycle_constrain}
    c(T_e) \leq c_{max}\,.
\end{align}
Similarly one introduces a \emph{release} cycle constraint.

Most of the scope of this paper is not dealing with non-local constraints. However we will return to the cycle constraint below in the section~\ref{sec:cycle_constraint}.

\section{Solution of deterministic problem}
\label{sec:Deterministic_problem}

\subsection{Penalty functions}
\label{sec:2.1}

To restrict $q(t)$ from going beyond the range $[Q_{min},Q_{max}]$, we can introduce a parametrised penalty function $-\phi[q(t),N_\phi]$ and add it to the unmodified Lagrangian~(\ref{eq:unperturbed_lagrangian})
\begin{align}
    L = L_0 -\phi(q) = -\dot qz - \gamma(\dot q) - \phi(q).
\end{align}

The penalty function is any smooth function, which is in the limit $N_\phi\to \infty$ approaches zero within the interval $(Q_{min},Q_{max})$ and positive infinity otherwise. A particular example of this function could be
\[   
    \phi = \Big[a(q(t)-b)\Big]^{2N_\phi}\,, \qquad \text{with} \quad
    a = \frac{2}{Q_{max}-Q_{min}}\,,\quad
    b = \frac{Q_{max} + Q_{min}}{2}\,.
\]

A similar penalty function $-\psi[\dot q(t),N_\psi]$ can be introduced to restrict~$\dot q(t)$ from going beyond the interval $[r_{min}, r_{max}]$. In the limit $N_\psi\to\infty$ it approaches zero within $(r_{min}, r_{max})$ and plus infinity otherwise. The modified Lagrangian becomes
\begin{align}
    \label{eq:perturbed_lagrangian}
    L = L_0 - \phi(q) - \psi(\dot q) = - \Big[ \dot q F + \phi(q) + \psi(\dot q) + \gamma(\dot q) \Big]\,.
\end{align}
It corresponds to the following modified action integral:
\begin{align}
    \label{eq:perturbed_action}
    S = - \int_0^{T_e} \Big[\dot q F + \phi(q) + \psi(\dot q) + \gamma(\dot q) \Big]\,dt\,.
\end{align}

\subsection{Euler-Lagrange equation}

As we know from the variational analysis, the extremal trajectory of the integral $\int L(q, \dot q)\,dt$ satisfies the Euler-Lagrange equation
\begin{align}
    \frac{\p L(q, \dot q)}{\p q} = \frac{d}{dt}\frac{\p L(q,\dot q)}{\p \dot q}\,.
\end{align}

Thus, for the modified Lagrangian we obtain the following equation
\begin{align}
    \label{eq:EL}
    \phi'(q) = \frac{d}{dt}\Big[F + \psi'(\dot q) + \gamma'(\dot q) \Big]\,,
\end{align}
which needs to be solved in the limit $N_\phi\to\infty,N_\psi\to\infty$.

There are two types of solution of the latter equation. One is obtained on the interval where the trajectory remains strictly within the boundaries
    \[ q(t) \in (Q_{min}, Q_{max} ).\]
The second solution type is, when the trajectory lies on the boundary
    \[ q(t) \equiv Q_{min}(t), \quad \text{ or } \quad q(t) \equiv Q_{max}(t)\,.\]
The general solution consists of pieces of the solutions of the types one and two. Let us consider each of these types separately.

\subsection{Solution within the boundaries}

First we consider the solution in the interval, where the trajectory does not touch the boundary. In the limit $N_\phi\to\infty$ the l.h.s. of Eq.~(\ref{eq:EL}) vanishes. We thus have
\begin{align}
    \label{eq:within_boundaries}
    F(t) + \psi'(\dot q) + \gamma'(\dot q) = C\,,
\end{align}
where $C$ is \textit{constant} for the whole period of time, where the solution does not touch the boundary.

Let us consider one particular example of the operating cost function. It is rather typical for a gas storage facility that the operating cost is proportional to the amount of the gas released or injected into the storage. Again we parametrise it with $N_\gamma$ so that for finite $N_\gamma$ the function $\gamma(\dot q)$ is smooth, and in the limit $N_\gamma\to \infty$ it is piecewise linear
\begin{align}
    \gamma(\dot q) = \left\{
        \begin{array}{cc}
            \gamma_{inj} \, \dot q\,, \quad& \text{for } \dot q > 0\,;\\
            -\gamma_{rel} \, \dot q\,, \quad& \text{for } \dot q < 0\,;
        \end{array}
    \right.
\end{align}
where $\gamma_{inj} > 0$ and $\gamma_{rel}>0$.

\begin{figure}[htb!]
  \begin{center}
    \includegraphics[angle= 0, width=0.5\columnwidth] {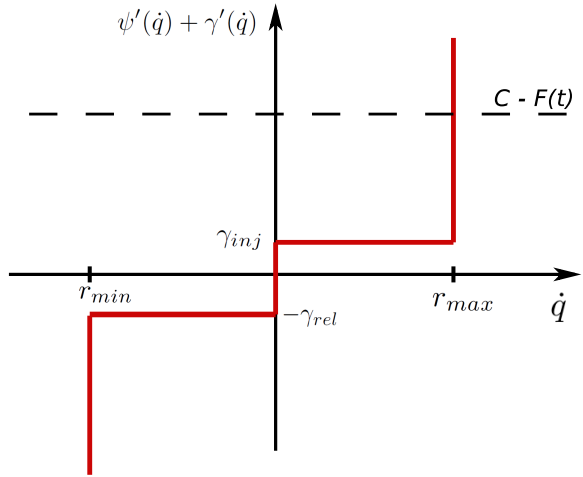}
    \caption{Graphical illustration of Eq.~(\ref{eq:within_boundaries}).}
    \label{fig:1}
  \end{center}
\end{figure}

The solution of Eq.~(\ref{eq:within_boundaries}) is now straightforward to find graphically (see Fig.~\ref{fig:1}). In case $r_{min}<0, r_{max}>0$ we obtain:
\begin{align}
    \label{eq:solution}
    \dot q(t) = \left\{
        \begin{array}{ll}
            r_{min}(t)\,, \qquad &  F(t) > C+\gamma_{rel}              \,; \\
            0\,,                 &  C-\gamma_{inj} < F(t) < C+\gamma_{rel} \,; \\
            r_{max}(t)\,, \qquad &  F(t) < C - \gamma_{inj}            \,;
        \end{array}
    \right.
\end{align}
We see that the value $C$ can be interpreted as a trigger price. If the price $F(t)$ is within the interval $[C-\gamma_{inj},\, C+\gamma_{rel}]$, the extremal trajectory is constant: $\dot q = 0$. If the price is below this interval, the volume $q(t)$ is increasing at maximal rate: $\dot q(t) = r_{max}(t)$. If the price is above the interval, the volume is decreasing at maximum rate: $\dot q(t) = r_{min}(t)$.

\subsection{Solution on the boundary}

We consider only a boundary $Q_{min}(t)$ and $Q_{max}(t)$, that moves slower than the limit rate, i.e. $r_{min}\leq \dot Q_{min} \leq r_{max}$ and $r_{min}\leq \dot Q_{max} \leq r_{max}$. If the boundary moves faster, then no trajectory can lie on it.

On the boundary the function $\psi'(\dot q)$ vanishes, and the function $\phi'(q)$ takes a finite value (positive if $q = Q_{max}$ or negative if $q = Q_{min}$). Eq.~(\ref{eq:EL}) on the boundary becomes
\begin{align}
    \label{eq:on_the_boundary}
    \phi'(q) = \frac{d}{dt} \Big[F(t)+\gamma'(\dot q)\Big]\,;
\end{align}

There is another view on the solution on the boundary. We may consider our problem as a general variational problem with spacial boundary condition. A path lying on the boundary can not be varied and hence be a part of extremal. But a solution, that maximises the functional, can consist of parts lying on the boundary and \emph{extremals} -- pieces of trajectory satisfying the extremal condition (i.e., vanishing first variation).

\subsection{Approaching the boundary}

An important statement can be made about the connection point between extremal and the boundary.  From the variational analysis it is known that an extremal touches the boundary smoothly.

The extremal solution of our variational problem for large but finite $N_{\phi,\psi,\gamma}$ is a smooth function and must satisfy the condition of smooth connection between extremal and boundary.
Let us consider a simplified problem (a generalisation is straightforward), where we suppose a constant boundary condition, and zero operating cost:
    \[ Q_{max} = const,\quad Q_{min} = const,\quad  \gamma(\dot q)\equiv 0\,.\]
In this case the extremal trajectory should touch the boundary at zero slope:
    \[\dot q(t^*) = 0\,, \]
where $t^*$ is the connection time. From Eq.~(\ref{eq:solution}) in the limit $\gamma_{rel}\to 0$ and $\gamma_{inj}\to 0$ we conclude that
\begin{align}
    F(t^*) = C
\end{align}
Taking the operating cost again into consideration leads to the following cases:
\begin{enumerate}
    \item The trajectory touches the lower boundary from the left:
        \begin{align}
            \label{eq:bc1}
            F(t^*) = C + \gamma_{rel}\,; \quad \dot F(t^*) \leq 0\,.
        \end{align}
    \item The trajectory touches the upper boundary from the left:
        \begin{align}
            \label{eq:bc2}
            F(t^*) = C - \gamma_{inj}\,; \quad \dot F(t^*) \geq 0\,.
        \end{align}
    \item The trajectory touches the lower boundary from the right:
        \begin{align}
            \label{eq:bc3}
            F(t^*) = C - \gamma_{inj}\,; \quad \dot F(t^*) \leq 0\,.
        \end{align}
    \item The trajectory touches the upper boundary from the right:
        \begin{align}
            \label{eq:bc4}
            F(t^*) = C + \gamma_{rel}\,; \quad \dot F(t^*) \geq 0\,.
        \end{align}
\end{enumerate}
These conditions must be satisfied for the time interval $t\in(0,T_e)$. The extremal trajectory on the times $t=0$ and $t=T_e$ does not have to satisfy these conditions.

The conditions (\ref{eq:bc1}-\ref{eq:bc4}) can be derived from a rule of thumb: if at the moment $t^*$ of boundary touch the boundary was virtually not there, the trajectory had to switch the mode anyway (like from ``full release'' to ``do nothing'' in case~(\ref{eq:bc1}), et.c.). The requirement to the price curve derivative follows from the simple observation: when the trajectory approaches the boundary, the forward curve should enter the dead zone, and when the trajectory leaves the boundary, the forward curve should exit the dead zone.

In the limit $N_{\phi,\psi,\gamma} \to \infty$ the solution may be non smooth any more. In particular the connection point $q(t^*)$ will be a point of the slope discontinuity.

It is worth emphasising that the constant $C$ may be different for different extremal pieces of trajectory separated by the boundary touch.

\subsection{Conclusion}

Eq.~(\ref{eq:solution}) together with boundary conditions (\ref{eq:bc1}-\ref{eq:bc4}) and condition on the start and end volume
\begin{align}
    \label{eq:init_term_cond}
    q(0) = Q_{start}\,;\qquad q(T_e) = Q_{end} = Q_{start} + \int_0^{T_e} \dot q \, dt
\end{align}
provides an implicit solution for the problem. We summarise some properties of the solution:
\begin{enumerate}
    \item The optimal exercise strategy is always bang-bang. If expressed in ``industrial'' language, the compressor either stands or works with a full power.
    \item For each piece of trajectory, separated from others with a boundary touch, there is a trigger price $C$ and a ``dead zone'' $[C-\gamma_{inj},\, C+\gamma_{rel}]$. The volume in the storage is increasing at maximum rate if the price is below the zone, decreasing at maximum rate if the price is above the zone, and is constant if the price is in the zone. The width of the zone is defined by the operating costs.
    \item If the trajectory touches the boundary in the time interval $t\in(0,T_e)$, it must satisfy one of the boundary conditions~(\ref{eq:bc1}-\ref{eq:bc4}).
    \item Generally a trajectory satisfying all the previous conditions is not unique. Any solution must obey all these conditions, but not any function obeying all conditions is a solution.
\end{enumerate}

The solution found in the previous section maximises the modified target functional $S[q(t)]$. Let $\bar q(t)$ be the obtained solution. Now the optimal path $\bar q(t)$ can be substituted to the unmodified functional $S_0[\bar q(t)]$ to give the value of option. A natural question is whether the trajectory $\bar q(t)$, which is optimal for modified functional, is still optimal for the unmodified one.

The optimal trajectory $\bar q$ belongs to the broad class of all locally integrable functions $\mathcal{L}_1$. Let us introduce a class of trajectories $\mathcal{K}\subset\mathcal{L}_1$ satisfying all the constraints. By definition penalty functions vanish on $\mathcal{K}$:
\[\phi(q)\equiv 0\,;\quad \psi(\dot q)\equiv 0\,,\quad \text{iff}\quad q\in\mathcal{K}\,.\]
Obviously the values of modified and unmodified functionals coincide on $\mathcal{K}$:
\[ S[q] = S_0[q]\,,\quad \forall q\in\mathcal{K}\,. \]

Next we show that the optimal trajectory $\bar q$ belongs to the class $\mathcal{K}$ and hence satisfies the constraints. Indeed, generally on some parts of the maximal trajectory $\psi'(\dot{\bar q}) \not = 0$  and $\phi'(\bar q) \not = 0$. According to Eqs.(\ref{eq:within_boundaries}) and (\ref{eq:on_the_boundary}) they take  finite values. One can show (we leave it without proof) that in the limit $N_{\phi,\psi}\to\infty$ everywhere where $\phi'$ (or $\psi'$) take finite value, the function $\phi$ (or $\psi$) vanishes. We conclude that on the extremal trajectory
\begin{align}
    \label{eq:zero}
    \phi(\bar q) \equiv 0\,;\qquad
    \psi(\dot{\bar q}) \equiv 0\,.
\end{align}
and hence $\bar q \in\mathcal{K}$. Form this it follows that the values of modified and unmodified target functionals coincide on the maximal trajectory:
\[S[\bar q] = S_0[\bar q]\,.\]

Last step is to prove that the trajectory $\bar q$ delivers the maximum of the unmodified functional. It is obvious from the following consideration. The trajectory $\bar q$ maximises the modified functional on the space $\mathcal{L}_1$ and hence it will also maximise this functional on the smaller space~$\mathcal{K}\subset\mathcal{L}_1$. Since the values of both functionals coincide on~$\mathcal{K}$, the trajectory $\bar q$ will also maximise the unmodified functional $S_0$ on the space of all trajectories satisfying constraints.

From this point we drop the subscript ``0'' for the action integral, unless we want to emphasise the difference between modified and unmodified target functions. We also drop the ``bar'' for the optimal trajectory $\bar q(t)$ everywhere where it does not lead to ambiguity.

\subsection{Other constraint types}
\label{sec:other_constraint_types}

\subsubsection{Carry cost}

Some storage contracts may include a ``carry cost'', which is the cost of keeping the commodity in the storage. For instance an oil storage is subject to a heating cost (a fuel oil in a storage has to be kept warm at approximately $60^oC$).

The carry cost is specified as a time-dependent price $\gamma_c(t)$ per unit time per unit volume. Thus, the target function becomes
\begin{eqnarray}
    \label{eq:action_carry_cost}
    S_0 = -\int_0^{T_e} \Big[ \dot q\,F(t) + \gamma(\dot q) + \gamma_c(t)\,q(t) \Big]\,dt
\end{eqnarray}
Substituting the new Lagrangian into the Lagrange-Euler equation we obtain between boundaries:
\begin{align}
    \label{eq:within_boundaries_carry_cost}
    F(t) + \psi'(\dot q) + \gamma'(\dot q) = C(t)\,
\end{align}
where
\begin{align}
    \label{eq:trigger_price_carry_cost}
    \frac{dC(t)}{dt} = \gamma_c(t)\,.
\end{align}
Thus, the solution of the problem with the carry cost is absolutely the same as that without but the only difference: the trigger price is not a constant, rather a function of time satisfying Eq.~(\ref{eq:trigger_price_carry_cost}).

\subsubsection{Solution with free terminal condition}
\label{sec:free_vs_fixed_terminal_condition}

Sometimes the storage problem can be defined with a free terminal condition. For instance, one is allowed to leave an arbitrary amount of underlying in the storage, and for the remaining volume one gets a additional pay-off equivalent to selling that volume at some effective price $F_e$, usually referred to as a \emph{final unit price}. We define the target functional $\tilde S$ as
\begin{align}
    \label{eq:2.18}
    \tilde S[q(t)] = -\int_0^{T_e} [ \dot q(t)\,F(t) + \gamma(\dot q)]\,dt + q_{end}\,F_e\,.
\end{align}
The optimisation problem is now to find an optimal exercise trajectory $q(t)\,,\ q_{end} = q(t_{end}) = q(T_e)$, satisfying operational constraints, that would maximise the target functions~(\ref{eq:2.18}). It is easy to see that any extremal trajectory of~(\ref{eq:2.18}) is also extremal of~(\ref{eq:unperturbed_action}). Indeed, if the trajectory~$\bar q(t)$ is extremal of~(\ref{eq:2.18}), the variation of the target functional vanishes on any small variation $\delta \bar q(t)$. Since there is no terminal boundary condition, the variation $\delta \bar q(t)$ may not vanish at the end point. But if the target function does not change on the whole class of allowed trajectories $q(t) = \bar q(t) + \delta\bar q(t)$, it also does not change on the sub-class of trajectories $q(t) = \bar q(t) + \delta \hat q(t)$ with fixed terminal value ($\delta \hat q(T_e) = 0$).

We conclude that to find a maximising trajectory for the functional~(\ref{eq:2.18}), we need first to find a solution for the functional~(\ref{eq:unperturbed_action}) with fixed terminal conditions. Then, considering this solution as a function of terminal state~$q_{end}$, we need to find a maximum of~(\ref{eq:2.18}) as a plain function of~$q_{end}$.

As will be shown in the Sec.~\ref{sec:3}, if the target functional~(\ref{eq:unperturbed_action}) is considered as a function of the terminal state~$q_{end}$, then its derivative with respect to $q_{end}$ is
\[
    \frac{\p S_0}{\p q_{end}} = -C\,.
\]
Thus, the target functional~(\ref{eq:2.18}), considered as a function of $q_{end}$, has a derivative
\begin{align}
    \label{eq:4.21}
    \frac{\p \tilde S}{\p q_{end}} = F_e - C \,,
\end{align}
where $C$ is the trigger price of the last part of trajectory. We conclude that the condition for a target functional with free terminal state to have maximum is
\begin{align}
    \label{eq:4.22}
    C = F_e\,.
\end{align}

In practice the storage option with free terminal state has two possibilities. One possibility is that the terminal state lies on the boundary $Q_{min}$ or $Q_{max}$. In this case the problem has a solution with fixed end. Another possibility is that the trajectory ends between boundaries $Q_{min} < q(T_e) < Q_{max}$. In this case the condition~(\ref{eq:4.22}) must be fulfilled.

If the storage option has no final unit price (price for the remaining volume), it is most likely to terminate with the lowest possible volume. Indeed, as we already know, the derivative of the storage option value with respect to the terminal state equals $-C$. If we neglect the operating costs, the trigger price can not be smaller than the smallest gas price. Hence, if the gas price is positive, then the derivative of the option value with respect to the terminal level is negative, and thus the optimal trajectory must terminate at the lowest possible level. Consequently the storage option can be considered as a problem with fixed terminal state (although it may not be fixed according to the contract).

An opposite example is a swing option. A typical swing option is a contract between a gas supplier and a trading company. The trader buys gas at some predefined price from the supplier and sells it on the market at the current market price. The spread between the market and supplier price becomes the effective gas price for the trader. The swing option allows the trader to take the gas from the supplier according to some flexible scheme (i.e. to decide, when to take and how much within certain constraints). Thus, the swing option can be formulated in terms of a storage option. However unlike the storage option, the effective price (the spread) in the case of swing option can be both positive and negative. Depending on the actual prices and on the contract constraints, it may happen that the terminal state does not lie on either lower or upper boundaries. In this case we deal with the problem with the free terminal condition. For this problem the trigger level equals zero
\begin{align}
    \label{eq:4.23}
    C = 0\,,
\end{align}
since there is no equivalent for the final unit price in the swing contract.

\subsubsection{Cycle constraint}
\label{sec:cycle_constraint}

The cycle constraint~(\ref{eq:cycle_constrain}) is formulated as a maximum allowed injected volume during the operation period $t\in[0,T_e]$. Instead of considering it in the form of inequality we can proceed as follows. We find a solution in two steps. On the first step we solve the problem without the cycle constraint and calculate the cycle variable on the optimal trajectory. If the cycle variable is below the threshold $c_{max}$, the obtained solution satisfies the cycle constraint. However if the solution violates the cycle constraint, we can reformulate the problem, imposing the ``biting'' cycle constraint
\begin{align}
    \label{eq:cycle_constrain_eq}
    c(T_e) = \int_0^{T_e} \dot q(t)\, \theta( \dot q(t) )\,dt = c_{max}
\end{align}
This condition allows to formulate our problem in terms of conditional variational extremum. It is solved by means of Lagrange multiplier. Namely we introduce a modified Lagrangian
\begin{align}
    \label{eq:modif_lagr_cycle}
    L = - \Big[ \dot q F + \phi(q) + \psi(\dot q) + \gamma(\dot q) + \lambda\, \dot q(t)\, \theta( \dot q(t) ) \Big]\,,
\end{align}
where $\lambda$ is an independent variable (Lagrange multiplier). The solution of the latter problem will contain the undefined coefficient $\lambda$, which has to be found from the additional equation~(\ref{eq:cycle_constrain_eq}).

Within the boundaries the Lagrange-Euler equation for the constrained Lagrangian becomes (compare to Eq.~(\ref{eq:within_boundaries}))
\begin{align}
    C = F(t) + \psi'(\dot q) + \gamma'(\dot q)  + \lambda \,\Big(\theta(\dot q) + \dot q\,\delta(\dot q) \Big) =
    F(t) + \psi'(\dot q) + \gamma'(\dot q)  + \lambda \,\theta(\dot q) \,.
\end{align}
Solving this equation we find that the solution is similar to that without cycle constraint, but has the ``dead zone'' by $\lambda$ wider. Thus, the cycle constraint is equivalent to some additional operating costs. One can find a fictive additional release and/or injection costs (same for the whole trajectory), such that if added to the storage without cycle constraint, the resulting trajectory would be extremal for the Lagrangian with imposed cycle constraint. It plays no role if the additional cost is for injection, release or both, since it is only the width of the dead zone which matters. The only exception from this rule is if the optimal trajectory does not inject or release enough volume. For instance if the trajectory never releases gas, the additional release cost will not effect the trajectory and will not help to respect the intake cycle constraint. Practically one should add injection costs to fulfil the intake cycle constraint, and release costs to fulfil the release cycle constraint.

Generally the solution between boundary touches contains two free parameters -- trigger price $C$ and cycle Lagrange multiplier $\lambda$. They can be equivalently expressed by upper and lower bounds of the dead zone. These parameters have to be found in such a way that the obtained solution satisfies boundary touch conditions~(\ref{eq:bc1}-\ref{eq:bc4}), initial and terminal conditions~(\ref{eq:init_term_cond}) and the cycle constraint~(\ref{eq:cycle_constrain_eq}).

\subsubsection{Volume dependent injection/release rates}
\label{sec:volume_dependend_rates}

If the maximal injection/release rates are volume dependent, the penalty function $\psi$ becomes an explicit function of volume: $\psi = \psi(q,\dot q)$. In this case the Euler-Lagrange equation within the boundaries becomes
\begin{align}
    \label{eq:4.25}
    \p_q \,\psi(q,\dot q)  = \frac{d}{dt}\Big[F + \p_{\dot q}\,\psi(q,\dot q) + \gamma'(\dot q) \Big]\,.
\end{align}
The l.h.s. of this equation is an integrable implicit function of time. Designating
\[ \frac{dC(t)}{dt} = \p_q \,\psi(q(\tau),\dot q(\tau)) \,;\qquad C(t) = C_0 + \int_0^t  \p_q \,\psi(q(\tau),\dot q(\tau)) \,d\tau \]
we obtain the same solution as in the case of volume independent rates with the difference that $C$ is not constant any longer, rather a function of time. We conclude that the optimal exercise strategy still preserves the ``bang-bang'' property. However is does not posses the constant trigger level. The function $C(t)$ can be reverse engineered from the optimal trajectory $q(t)$, provided the latter is been found.

Let $q(t)$ be the optimal trajectory. On this trajectory the derivative $\p_{\dot q}\,\psi(q,\dot q)$ takes a finite value, which can be found from the equation
\[ \p_{\dot q}\,\psi(q(t),\dot q(t)) = C(t) -F(t) - \gamma'(\dot q(t))\,.  \]
We can also find the relation between the derivatives $\p_q\,\psi(q,\dot q)$ and $\p_{\dot q}\,\psi(q,\dot q)$. Indeed, if the trajectory $q(t)$ lies for instance on the maximum injection rate boundary $\dot q = r_{max}(q)$, then the relation between $dq$ and $d\dot q$ is simply $d \dot q = r'_{max}(q)\,dq$. It's easy to show that the gradient $\{\p_q\,\psi(q,\dot q),\p_{\dot q}\,\psi(q,\dot q)\}$ must be orthogonal to the vector $\{dq, d\dot q\}$. Hence
\[
    \p_q\,\psi(q,\dot q) = r'_{max}(q)\,\p_{\dot q}\,\psi(q,\dot q) \,.
\]
We thus obtain the differential equation
\[
    \frac{dC(t)}{dt} = r'(q(t)) \Big( C(t) -F(t) - \gamma'(\dot q(t)) \Big)\,. \qquad
    r'(q(t)) = \left.\frac{dr(q)}{dq}\right|_{q = q(t)}\,.
 \]
Here $r(q) = r_{max}$ must be used for ``injection'' and $r_{min}$ for ``release'' parts of the trajectory. This equation allows to find the function $C(t)$ provided the optimal trajectory $q(t)$ is already known.

\subsection{Dependency on initial and terminal conditions}
\label{sec:3}


The optimal path $q(t)$ and the target functional $S[q(t)]$ are calculated for fixed terminal point $\{T_e,Q_{end}\}$. If this point is slightly shifted in either direction -- time or volume -- the optimal path becomes different, and the target function as well. In this sense the target function can be viewed as a function of terminal point $S = S(q,t)$. In this section we are interested in the properties of this function.

From the standard variational analysis we have:
\begin{align}
    \frac{\partial S}{\partial q} = \frac{\partial L}{\partial \dot q}\,; \qquad
    \frac{\partial S}{\partial t} = L - \dot q \frac{\partial L}{\partial \dot q}\,.
\end{align}
We apply these equations to the perturbed action $S[q(t)]$. Making use of Eq.~(\ref{eq:zero}) we will find the derivatives of the unperturbed action $S_0[q(t)]$.

\subsubsection{Volume derivative}
\label{sec:spacial_derivative}

Using the definition (\ref{eq:perturbed_lagrangian}) of the perturbed Lagrangian we find
\[
    \frac{\partial S}{\partial q} = - \Big[ F + \psi'(\dot q) + \gamma'(\dot q) \Big]\,.
\]
The spacial derivative makes sense only within the boundaries. Making use of Eq.~(\ref{eq:within_boundaries}) and replacing $S$ with $S_0$ we finally find
\begin{align}
    \label{eq:spacial_derivative}
    \frac{\partial S_0}{\partial q_{end}} = \frac{\partial S}{\partial q_{end}} = - C\,,
\end{align}
where $C$ is the trigger level of the last part of the trajectory. If $C$ is positive, one can draw a conclusion, that the smaller the terminal state $q(T_e)$, the bigger is the target functional $S_0(T_e)$.

Derivative with respect to the initial condition $q_{start}$ is given by inverting the sign:
\begin{align}
    \frac{\partial S_0}{\partial q_{start}} = \frac{\partial S}{\partial q_{start}} = C\,,
\end{align}
where $C$ is the trigger level of the initial part of the trajectory.

From this an important consequence follows, which we already mentioned in the section~\ref{sec:free_vs_fixed_terminal_condition}. If there is no final unit price and if the prices $F(t)$ are strictly positive, so is the trigger price~$C$. As a result the option value has a negative derivative with respect to the terminal level. Hence the optimal trajectory will always take at the end the smallest possible value allowed by constraints. By other words, we know that the right end of the optimal trajectory will lie on the lower boundary. Thus, even if the contract allows the storage to end with some remaining volume inside (free terminal condition), the optimisation problem still can be considered as the one with fixed terminal condition.

This is not any longer the case in presence a positive final unit price is given or if the prices $F(t)$ can be negative. With the final unit price $F_e$ the option value derivative is given by Eq.~(\ref{eq:4.21}) and can be positive, negative or zero. Also extremely high injection/release costs may lead to vanishing trigger price and the option value derivative.

\subsubsection{Time derivative}

From definition we find
\begin{align}
    \frac{\partial S}{\partial t}& =
    \dot q\Big[\psi'(\dot q) + \gamma'(\dot q)\Big] -
        \phi(q) - \psi(\dot q) - \gamma(\dot q) = \nonumber \\
    & = \dot q\Big[\psi'(\dot q) + \gamma'(\dot q)\Big] - \gamma(\dot q)\,.
\end{align}
\textbf{Within the boundaries} we make use of Eq.~(\ref{eq:within_boundaries}). Replacing $S$ with $S_0$ one obtains
\begin{align}
    \frac{\partial S_0}{\partial t_{end}} = \frac{\partial S}{\partial t_{end}} = \dot q(t_{end}) \Big[C - F(t_{end}) \Big] - \gamma(\dot q)\,,
\end{align}
where $\dot q(t_{end})$ is the solution (\ref{eq:solution}) obtained for fixed terminal condition. In particular one can see that  $\partial S_0/\partial t_{end} \geq 0$ (provided $r_{min}<0, r_{max}>0$).

\textbf{On the boundary} in the special case $\dot Q_{max} = \dot Q_{min} = 0$ one obtains $\dot q= 0, \gamma(\dot q)= 0$ and
\begin{align}
    \frac{\partial S_0}{\partial t_{end}} = 0\,.
\end{align}
It is worth noting that the derivatives with respect to terminal (initial) boundary conditions is universal: it is independent on whether the optimal trajectory touches the boundary or not. To calculate the derivatives, one just needs to know the solution $q(t)$ and the trigger price $C$ at the end (beginning) of the trajectory. This also remains valid with imposed cycle constraint.

%
%
%
%

\section{Stochastic problem}
\label{sec:stochastic_problem}

The following section is devoted to the stochastic optimisation problem. As a starting point of the stochastic approach we take the solution of the corresponding intrinsic problem, which is supposed to be known. Under some very generic assumptions about the price process we derive a stochastic differential equation which governs the evolution of the storage value in the rolling intrinsic approximation. This allows us to find the expectation value of the terminal option value, and hence to estimate the time value of the storage option.

In real world the prices are not static, rather evolve with time in a stochastic way. Every time we observe a new forward curve on the market, we can solve a new intrinsic problem, obtaining a new solution. The forward price $F$, as well as the optimal intrinsic strategy $\dot {\bar q}$ and trigger price $C$ become not only functions of ``delivery'' (or ``maturity'') time $T$, but also functions of the ``current'' (or ``observation'') time~$t$:
\[
    C = C(t)\,;\quad F = F(t,T)\,;\quad \dot {\bar q} = \dot {\bar q}(t,T)\,.
\]
By other words, $F(t,T)$ is the forward price with delivery time $T$ observed on the market at time~$t$. $C(t)$ and $\dot {\bar q}(t,T)$ are trigger price and optimal intrinsic strategy, calculated at time $t$ according to current observed forward price curve $F(t,T)$. Throughout the following section we designate
\[ r(t, T) \equiv  \dot {\bar q}(t, T)\]
the optimal intrinsic exercise strategy on the observation time $t$.

By definition of the stochastic problem, the initial forward curve $F(0,T)$ is known. For any future observation time $t>0$ there is no deterministic forward curve. The evolution of the forward curve is supposed to satisfy some stochastic differential equation. It means that any possible curve $F(t,T)$ can be assigned a certain probability amplitude, and the solution of the problem has to be formulated in the probabilistic language.

\paragraph{Assumptions:} To make further analysis we make some simplifying assumptions. 
\begin{itemize}
    \item First we neglect the injection/release costs by setting  $\gamma\equiv 0$. 
    \item We also neglect the ``surface effects'' -- the possible influence of the boundary touch on the optimal trajectory and on the storage option value. Under the specified assumptions for every observation time $t$ there exists a single trigger price $C(t)$ for the whole trajectory, and the optimal intrinsic strategy can be presented in the form
\begin{align}
    \label{eq:solution_for_dot_q}
    r(t, T) = r_{min} + \Delta r\,\theta(C(t) - F(t,T))\,,
\end{align}
where $\Delta r = r_{max} - r_{min}$, $\theta(x)$ is the Heaviside function and $C(t)$ is the intrinsic trigger price at time $t$. We also introduce the following notation:
\begin{align}
    r^{(1)}(t,T) = \Delta r\,\delta(C(t)-F(t,T))\,;\qquad
    r^{(2)}(t,T) = \Delta r\,\delta'(C(t)-F(t,T))\,.
\end{align}
Here $\delta(x) = \theta'(x)$ is the Dirac delta function. The derivative is taken with respect to the argument of the $\delta$-function. Notice the dimensionality:
\begin{align}
    r = \left[\frac{Q}{t}\right]\,; \qquad
    r^{(1)} = \left[\frac{Q}{t\,\$}\right]\,; \qquad
    r^{(2)} = \left[\frac{Q}{t\,\$^2}\right]\,.
\end{align}

    By neglecting the ``surface effects'' we restrict our consideration to the so called non-flexible storages. The storage flexibility can be characterised by a number of times $N_c$ the storage can be filled and emptied completely during the storage life time (maximum number of cycles). The storage is usually called non-flexible if $N_c \lesssim 1$. For $N_c \gg 1$ the storage is referred to as flexible. Obviously, the surface effects have much bigger impact on the flexible storages.

    \item As an additional assumption we believe that the prices $F(t,T)$ are always strictly positive. This allows us to consider the optimisation problem as a problem with fixed initial and terminal conditions (see Sec.~{\ref{sec:spacial_derivative}}). This condition is applicable for the storage option and is not for the swing option. The swing option will be considered separately in the section~\ref{sec:swing_option}.
\end{itemize}

\subsection{Formulation of the stochastic optimisation problem}
\label{sec:stoch_opt_problem}

For the intrinsic problem one could determine an optimal exercise strategy $r(T)$ which has a meaning of amount (per unit time) of underlying which has to be injected/released from the storage and bought/sold on the market at the time~$T$. It makes no difference if this amount is sold on the spot (i.e. exactly on the time~$T$) or on the forward market prior the delivery time $T$, since the price of the underlying does not change.

This is not the case in the stochastic reality. Since the price for the delivery on time $T$ changes with the observation time $t$, it is essential to specify when exactly the underlying is traded. We thus distinguish between spot trades and forward trades. A spot trade is associated with immediate delivery, and the corresponding price is referred to as ``spot price'' $s(t) = F(t,T=t)$. A forward trade has a delivery in the future, and generally can be any kind of financial products, including forward contracts or options. The forward trades usually have a purpose of financial hedging, so we will refer to them as \emph{hedge trades}. For simplicity we only consider linear hedge products, i.e. forward contracts. Thus for the stochastic problem at every time $t$ we can introduce a trade profile $h(t,T)$, which has the following meaning:
\begin{itemize}
	\item The up-front value $h(t,t)$ is the exercise trade. The amount of underlying $h(t,t)\,dt$ has to be injected in the storage within the time $dt$, and the corresponding amount has to be purchased on the market. The purchased amount is a net amount purchased on the forward and spot markets for the delivery on $T=t$. We will refer to the exercise trade $h(t,t)$ as \emph{prompt exercise}.
    \item The volume $h(t,T)$ for $T>t$ is a hedge position. The value $h(t,T)\,dT$ has a meaning of the total volume purchased on the forward market (i.e. aggregated volume of all forward trades performed during the time period $[0,t]$) for the delivery period $[T,T+dT]$.
\end{itemize}

For the stochastic optimisation problem there exists no optimal exercise trajectory, which could be computed prior or during the contract time. Hence it is not possible do define a target functional in the form of integral~(\ref{eq:unperturbed_action}). Instead we have to formulate the aim of the stochastic optimisation problem as finding an optimal trading strategy $h(t,T)$, such that the expectation value of the total profit is maximised.

The trading profile can be used for exercise and hedge trades on the future market, giving rise to the cash flow at the observation time\footnote{In reality the exact timing of the cash flow depends on the type of financial instruments used for hedging, and can be spread between observation and delivery time. However it plays no role for our analysis, and we agree to associate the cash-flow from the forward deals with the trading (observation) time~$t$. Thus we agree that if the volume $V$ is purchased on the observation time $t$ with delivery on $T$, then the corresponding cash-flow $-V\,F(t,T)$ is associated with the observation time $t$.}. Since the hedge profile is different for every time moment~$t$, one has to update the hedge position continuously by buying and selling a proper amount of the underlying on the future and spot markets. Let $h(t,T)$ be the hedge position taken by the time~$t$. This also includes the exercise volume $h(t,t)\,dt$, which has to be delivered during the time~$dt$. After the infinitesimal time increment $dt$ we observe the new prices on the market, and find a new optimal hedge strategy $h(t+dt,T)$. The volume
\[ \delta h(t,T) = h(t+dt,T) - h(t,T)\]
has to be purchased on the market in order to update the hedge position. This volume includes the spot trade $\delta h(t,t)\,dt$ and hedge trades $\delta h(t,T)\,dT$ for $T>t$. Of the whole hedge profile $h(t,T)$ the only initial value 
\begin{align}
    \label{eq:hedge_profile_exercise_part}
    \left. h(t, T)\right|_{T=t}\,.
\end{align}
is associated with the ``physical'' delivery of the underlying, and the rest is covered by the forward contracts. 

Our aim now is to find for every observation time $t$ an optimal hedge profile $h(t,T)$ conditional on the market state $F(t,T)$ and current initial condition $q(t)$. In order to give a precise mathematical definition of the stochastic optimisation problem we need to define the stochastic profit function.

Let $h(0,T)$ be the initial hedge profile, calculated on the basis of the initial conditions $F(0,T)$ and $q(0) = Q_{start}$. The integral
\begin{align}
    \label{eq:4.2}
    P_0 = - \int_0^{T_e} h(0,T) \, F(0,T)\,dt\,.
\end{align}
gives the total cost of setting up the initial hedge profile. Let $h(t,T)$ be the hedge profile at time $t$. We introduce the profit function $P(t)$, which is defined as a cumulative cash flow from all hedge trades. The profit function satisfies the following equations:
\begin{align}
    \label{eq:5.5}
    &P(0) = P_0 = -\int_0^{T_e} h(0,T)\,F(0,T)\,dT \\
    \label{eq:5.6}
    &dP(t) = - \int_t^{T_e} \delta h(t,T)\,
    \Big( F(t,T) + \delta F(t,T) \Big)\,dT \,,
\end{align}
where $\delta h$ and $\delta F$ are the increments of the optimal hedge strategy and forward curve within a time-interval~$dt$. The stochastic optimisation problem can now be formulated as finding at every time $t$ an optimal hedge profile $h(t,T)$ such that the expectation value of the terminal profit $P(T_e)$ is maximised.

The defined profit function does not represent the total portfolio value, since it only reflects the cash-flow and does not take into account the value of the storage option itself. However this is the only part of the portfolio, which can be influenced by our trading strategy, and hence optimisation of the profit function is equivalent to the optimisation of the Profit and Loss.

Note that the hedge profile $h(t,T)$ for $T>t$ does not need to satisfy the storage constraints, since it is not associated with the physical delivery. The only physical part of the hedge profile is the prompt exercise trade $h(t,t)$ which has to obey the storage constraints.

Notice that the integral (\ref{eq:4.2}) coincides with the target functional~(\ref{eq:unperturbed_action}) (we have neglected the operating costs). As an example of a (non-optimal) trading strategy we can do the following. Let the initial hedge profile coincides with the initial intrinsic strategy
\[h(0,T) = r(0,T)\]
In this case the initial value of the profit function coincides with the intrinsic value $P(0) = P_0 = S_0$. If we decided not to change the hedge profile with the time\footnote{We can do so because the intrinsic profile satisfies the storage constraints, and hence we will be able to fulfil the contractual obligations.}
\[h(t,T) = h(0,T)\]
then the profit function would remain constant, giving rise to the terminal profit equal to the initial intrinsic value. Thus the stochastic profit is at least as big as the intrinsic value. In reality for every time $t$ one can almost always find a new hedge profile $h(t,T)$ such that the profit function increment $\delta P$ is positive. The increase of the terminal profit function in comparison with the intrinsic value is referred to as \emph{time value}.

Running ahead it is also interesting to notice that only the prompt exercise decision $h(t,t)$ is relevant for the expectation value of the profit. The rest of the hedge profile $h(t,T)$ for $T>t$ only effects the distribution of the terminal profit as a stochastic value. If our model of the stochastic price process was exact, and if we neglected the market frictions, then it would be theoretically possible to find such a hedge profile, which would make the distribution of the terminal profit function arbitrarily narrow.

\subsection{Rolling intrinsic strategy}

The optimal intrinsic solution at time $t$ is a function $\dot q(T) = r(t,T)$, which maximises the deterministic functional
\begin{align}
    - \int_t^{T_e} r(t,T) \, F(t,T)\,dT
\end{align}
given the initial condition $q(t)$ and the market state $F(t,T)$ at the time $t$, and subject to the storage constraints.

If at every time moment $t$ the hedge strategy $h(t,T)$ coincides with the current optimal intrinsic exercise
\[ h(t,T) = r(t,T)\,,\]
then the strategy is called \emph{rolling intrinsic}.

In the Sec.~\ref{sec:4.6.4} we show that under very general assumptions the rolling intrinsic is very good approximation of the optimal stochastic exercise strategy. Based on this fact and taking into account that the hedge strategy does not effect the expected value of the profit function, one can conclude that if we follow the rolling intrinsic strategy, the expected value of the terminal profit must be equal the true option value. This also allows one to calculate the time value as an average cumulative cash flow from all hedge and exercise trades. In App.~\ref{ap:A2} we give a systematic proof of this fact.

\subsection{Forward curve evolution model}

Let $F(t,T)$ for each time $T$ be a zero mean stochastic process governed by the stochastic differential equation in Ito representation
\begin{align}
    \label{eq:generic_price_process}
    &\frac{dF(t,T)}{F(t,T)} = dM_t(T)\,,
\end{align}
where $M_t(T)$ is a martingale process with respect to the variable $t$ parametrised by the variable~$T$. In general for each future time $T$ the martingale processes $M_t(T)$ are different processes, which might or might not be correlated.

The correlation between processes $M(T)$ is given by
\begin{align*}
    & \la\, dM_t(T_1)\,dM_t(T_2)\,\ra = \sigma^2(t,T_1,T_2)\,dt\,;\qquad
    \la\, dM_{t_1}(T_1)\,dM_{t_2}(T_2)\,\ra = 0 \quad \text{for} \quad t_1 \neq t_2\,,
\end{align*}
where the function $\sigma(t,T_1,T_2)$ is supposed to be a decaying function of $|T_1-t|$, $|T_2-t|$, and $|T_2-T_1|$. In the stochastic calculus the product $dM_{t_1}(T_1)\,dM_{t_2}(T_2)$ can be considered as deterministic and requires no averaging.

Let us designate
\begin{align}
    L(t, T_1,T_2) =  dF(t, T_1)\,dF(t, T_2) = F(t,T_1)\,F(t,T_2)\,\sigma^2(t,T_1,T_2)\,dt\,.
\end{align}
Notice that $L$ is a stochastic variable, since it depends explicitly on the stochastic prices $F$.

We define the price process \emph{correlation function} as
\begin{align}
    \Lambda(t,T_1,T_2) = \frac{1}{dt}\,\la L(t,T_1,T_2) \ra_{F} = 
    \sigma^2(t,T_1,T_2) \,\la F(t,T_1)\,F(t,T_2) \ra\,,
\end{align}
where $\la\cdot\ra_{F}$ is an averaging over the stochastic prices ~$F$.

As a particular example we consider the following process
\begin{align}
    & dM_t(T) = \sigma_0\,e^{-\alpha(T-t)}dW_t(T)\,,\\
    & \la\,dM_t(T_1)\,dM_t(T_2)\,\ra =
    \sigma_0^2\,e^{-\alpha(T_1-t)}\,
    e^{-\alpha(T_2-t)}\,
    e^{-\beta\,|T_2-T_1|}\,dt\,.
\end{align}
where $W_t$ is a standard Brownian motion. The price process has the following correlation function
\begin{align}
    \Lambda(t, T_1,T_2) =
    \sigma_0^2\,e^{-\alpha(T_1-t)}\,
    e^{-\alpha(T_2-t)}\,
    e^{-\beta\,|T_2-T_1|}\, \la F(t,T_1)\,F(t,T_2) \ra\,.
\end{align}
Note that the case $\beta=0$ corresponds to the standard one-factor forward curve model. In this case the whole curve is driven by a single Wiener process $W_t$.

To find the correlation $\la F(t,T_1)\,F(t,T_2)\ra$ we need to integrate the price process~(\ref{eq:generic_price_process}). For one-factor model we obtain
\begin{align}
     &\la F(t,T_1)\,F(t,T_2)\ra  = F_0(T_1)\,F_0(T_2)\,l(t,T_1,T_2)\,, \qquad\text{where} \\
     &l(t,T_1,T_2)=  \exp\left[
        \frac{\sigma_0^2}{2\,\alpha}\, e^{-\alpha\,(\tau_1+\tau_2)} \left(1  - e^{-2\,\alpha\,t } \right)
        \right]\,; \quad \tau_1 = T_1-t\,;\quad \tau_2 = T_2-t\,. \nonumber
\end{align}
Note also that after a relaxation time $\approx 1/\alpha$ the normalised correlation function $l$ becomes stationary, i.e. depending only on the time differences $\tau_1$ and $\tau_2$.

It can be easily shown that for small volatility $\sigma_0^2\,t \ll 1$ the function $l(t,T_1,T_2)\approx 1$, and thus for one-factor model we get
\begin{align}
     \Lambda(t, T_1,T_2) \approx    
    \sigma_0^2\,e^{-\alpha( \tau_1 + \tau_2  )} \,F_0(T_1)\,F_0(T_2)\,.
\end{align}

\subsection{Constraint surface}

We apply a variational analysis technique to find the increments of the hedge position $\delta h(t,T)$ in the rolling intrinsic approximation, and to derive a stochastic differential equation for the storage option value evolution based on Eqs.~(\ref{eq:5.5}) and~(\ref{eq:5.6}).

When finding the functional derivatives, it is important to distinguish between the two following cases. Let ${\cal L}[\phi_1(t), \phi_2(t)]$ be a functional defined on the functions $\phi_1(t)$ and $\phi_2(t)$. If they are considered as independent variables, the corresponding functional derivatives
\[
    \frac{\delta{\cal L}}{\delta\phi_1(t)} \qquad \text{and} \qquad \frac{\delta{\cal L}}{\delta\phi_2(t)}
\]
are analogous to the partial derivatives of an ordinary function of two variables.

On the other hand, if the variables $\phi_1$ and $\phi_2$ are dependent, then the variation
\[
    \delta{\cal L} = \frac{\delta{\cal L}}{\delta\phi_1(t)}\,\delta\phi_1(t) +
    \frac{\delta{\cal L}}{\delta\phi_2(t)}\,\delta\phi_2(t)\,; \qquad
    \delta\phi_2(t) = f(\delta\phi_1(t))
\]
is analogous to a total derivative of an ordinary function.

Let us consider a functional ${\cal L}$ defined on the variables $t,C(t),F(t,T)$. If we consider a time evolution of the forward curve $F(t,T)$, then all three increments $dt, \delta C(t)$ and $\delta F(t,T)$ become dependent. The relation between these increments follows from the requirement that the terminal level must be preserved:
\begin{align}
    & \delta Q_{end} = 0\,; \qquad \text{where} \nonumber \\
    \label{eq:constraint}
    & Q_{end} = Q(t) + \int_t^{T_e} r(t,T)\,dT \,;
\end{align}
The equation $\delta Q_{end} = 0$ for every time $t$ defines a surface in the space of $F(t,T)$ and $C(t)$, which restricts the possible simultaneous variations of the trigger price and forward curve. We say that the equation $\delta Q_{end} = 0$ defines a \emph{constraint surface}. If we find a variation $\delta {\cal L}$ which takes into account the relation between $\delta t$, $\delta C(t)$ and $\delta F(t, T)$, we refer to this variation as a \emph{variation on the constraint surface}.

Since the time-dependent increment of the forward curve $\delta F$ is driven by a Wiener-type stochastic equation of motion, we will treat this increment in a stochastic sense. In the framework of the stochastic calculus the square increment $\delta F^2$ can not be neglected if compared with the first order variation, since
\[ \delta F^2 \sim dt\,.\]
Hence the time-dependent variational derivatives must contain the second order terms alongside with the first order terms.

The details of derivation can be found in the Appendix~\ref{ap:Contraint_surface}. After some algebra we finally obtain
\begin{align}
    \label{eq:dC_2_order}
    &\delta C = \frac{1}{K}
        \left(  \int r^{(1)} \,\delta F\,dT  -
                \frac12 \int r^{(2)}\,\delta F^2 \,dT +
                \frac{1}{K}\iint r^{(2)}_u\,r^{(1)}_v\,L_{uv} -
                \frac{M}{2\,K^2}\iint r^{(1)}_u\,r^{(1)}_v\,L_{uv}
          \right)\,; \\
    & \delta C^2 =\frac{1}{K^2}\iint r^{(1)}_u\,r^{(1)}_v\,L_{uv}\,,
\end{align}
where we have used the notation
\begin{align*}
    & r^{(1)}(t,T)   = \Delta r\, \delta(C(t)-F(t,T)) \,;  && r^{(2)}(t,T) = \Delta r\, \delta'(C(t)-F(t,T))\,; \\
    & K(t) = \int_t^{T_e} r^{(1)}(t,T)\,dT\,;              && M(t) = \int_t^{T_e} r^{(2)}(t,T)\,dT\,; \\
    & L_{uv} = \delta F(u) \, \delta F(v)\,. 
\end{align*}
In particular with this notation we can write
\begin{align}
      \left[\int r^{(1)}\,\delta F\,dt\right]^2 =
      \iint r^{(1)}(u)\,r^{(1)}(v)\, \delta F(u) \, \delta F(v) \,du\,dv =
      \iint r^{(1)}_u\,r^{(1)}_v\,L_{uv}\,.
\end{align}
Eq.~(\ref{eq:dC_2_order}) defines the constraint surface. It relates the increments $dt, \delta C(t)$ and $\delta F(t,T)$. Note that according to the forward curve evolution model $L_{uv}\sim\delta t$.

The first order variation of the trigger price on the constraint surface is then given by
\begin{align}
    \label{eq:dC_1_order}
    & dC(t) = \frac{1}{K(t)} \int_t^{T_e} r^{(1)}(t,T) \,\delta F(T)\,dT \,.
\end{align}

\subsection{Stochastic trigger price}
\label{sec:4.6.4}

In the deterministic formulation of the storage problem, for each predefined forward curve $F(t,T)$ we can find an optimal exercise trajectory $\dot{\bar q}(T) = r(t,T)$. The value of the target function $S$ on the optimal trajectory at time $t=0$ is called ``intrinsic value''. If the prices are stochastic, there exist no optimal exercise trajectory, and the solution of the stochastic problem can only provide an optimal prompt exercise $r(t,t)$ at every time $t$.

In this section we show that the optimal stochastic exercise is bang-bang, and that for every time moment $t$ there exist a stochastic trigger price $C_{st}(t)$. We also show that the intrinsic trigger price is a good approximation for the stochastic one, and derive the conditions for this approximation.

Let $F(t,T)$ be the forward curve observed on the time $t$. We define a \emph{spot price} process $s(t)$ as
\begin{align}
    s(t) := F(t,t).
\end{align}
Since the forward price curve evolves with time, the spot price $s(t) = F(t,t)$ generally does not follow the original forward curve ($F(t,t) \neq F(0,t)$ almost everywhere). The spot process $s(t)$ is uniquely defined by the forward price process, and thus to each possible spot price curve $s(t)$ one can assign an amplitude or, in case of discrete price and time space, a probability $p$.

Let us consider now a set of all possible spot price curves $s_i(t)$ each having corresponding probability $p_i$. Considering the spot price curve $s_i(t)$ as a deterministic function, we can solve a static optimisation problem for the target functional
\[S_i = -\int_0^{T_e} \dot q(t)\,s_i(t)\,dt\,.\]
Thus, for every spot price curve $s_i(t)$ we can find a corresponding trigger price $C_i$, exercise strategy $r_i(t)$ and intrinsic value $S_i$. On each spot price path the trigger price is different, and the exercise strategy $r_i(t)$ is optimal only for that path.

Now we are looking for an optimal prompt exercise decision $\dot q(0)$ for the stochastic problem. Since we can only make one exercise decision at a time, this exercise decision will not be optimal for some spot price paths. If on the $i$th spot price path the optimal exercise is $r_i(0)\neq\dot q(0)$, the suboptimal exercise decision will lead to a loss of value on that path. We need to make such a choice of $\dot q$ that would minimise the losses on each path, for which the decision is not optimal.

During the time interval $dt$ the volume increment in the storage is given by $dq = \dot q\,dt$. Thus, on the $i$th spot price path the volume in the storage is by $(\dot q(0) - r_i(0))\,dt$ bigger than it had to be, if the exercise was optimal for that path. It leads to the change of the value on the $i$th path by
\begin{align}
    \label{eq:4.13}
    \delta S_i = \Big( C_i - F(0,0) \Big)\Big( \dot q(0) - r_i(0)\Big)dt
\end{align}
Indeed, the change of the value has two reasons -- additional expenses $\delta S_{i1}$ of buying an additional volume, and increased value $\delta S_{i2}$ due to change in the actual storage volume. The price of additional volume is
\[\delta S_{i1} = - F(0,0)\, \Big( \dot q(0) - r_i(0)\Big)\,dt.\]
The change of the value due to change of initial volume follows from the formula $\p S/ \p q_{start} = C$ (see Sec.~\ref{sec:3}):
\[\delta S_{i2} = C_i \,\Big( \dot q(0) - r_i(0)\Big)\,dt\,.\]
Combining the latter two equations we obtain Eq.~(\ref{eq:4.13}). Now we average $\delta S$ over all paths:
\begin{align}
    \la \delta S \ra = \sum_i p_i \Big( C_i - F(0,0) \Big)
        \Big( \dot q(0) - r_i(0)\Big)dt\,,
\end{align}
where $p_i$ is the probability of the $i$th path. The optimal exercise decision $\dot q(0)$ should maximise the expected change of option value $\la \delta S \ra$. Deriving $\la \delta S \ra$ with respect to $\dot q(0)$ we obtain
\begin{align}
    \frac{\p \la \delta S \ra}{\p \dot q(0)} = \sum_i p_i\,\Big( C_i - F(0,0) \Big)dt =
        \Big( \la C \ra  - F(0,0)\Big) \,dt\,,
\end{align}
where $\la C \ra = \sum_i\,p_i\,C_i$ is the trigger price averaged over all possible spot price processes. If the spot price $F(0,0)$ is bigger than the average trigger price $\la C\ra$, then the derivative is negative, and the optimal stochastic exercise is the smallest possible allowed by constraints. Similarly if $F(0,0)$ is smaller than the average trigger price, then the optimal exercise is the biggest possible allowed by constraints. Thus, the optimal prompt stochastic exercise is
\begin{align}
    \dot q(0) = \left\{
        \begin{array}{ll}
            r_{min}(t)\,, \qquad &  F(t) > \la C \ra             \,; \\
            r_{max}(t)\,, \qquad &  F(t) < \la C \ra            \,;
        \end{array}
    \right.
\end{align}
We see that the optimal stochastic exercise is bang-bang, and that the expected value of the intrinsic trigger price $\la C \ra$ can be interpreted as the stochastic trigger price $C_{st}$:
\begin{align}
    \label{eq:stochastic_trigger_price_1}
    C_{st} = \la C \ra\,.
\end{align}

Next we can show that under some assumptions the stochastic trigger price in the leading order of Taylor expansion equals the intrinsic trigger price. Indeed, let us designate $s_0(t)$ that spot price path which coincides with the initial forward curve at time $t= 0$:
\[s_0(t) = F(0,t)\,.\]
The trigger price $C_0$ is then the intrinsic trigger price, calculated at time $t=0$ on the basis of the forward curve $F(0,t)$. All other paths $s_i(t)$ are different, and we designate
\[\delta s_i(t) = s_i(t) - s_0(t)\,.\]
This difference is zero at the beginning of time period: $\delta s_i(0) = 0$ since all paths of the spot price process start in the same point $s_i(0) = F(0,0)$. At the end of the period the difference $\delta s_i(T_e)$ may be arbitrary large, but the majority of paths stay within the range
\[\sqrt{\lla \delta s^2(T_e) \rra}\]
If $\delta s_i(t)$ remains small, then the difference in the trigger level $\delta C_i = C_i-C_0$ can be calculated by means of the first order Taylor expansion~(\ref{eq:dC_1_order}):
\[
    \delta C_i \approx \frac{\Delta r}{K(0)} \int_0^{T_e} \delta(C_0-s_0(t))\,\delta s_i(t)\,dt\,.
\]
Since the forward price process is a martingale, we conclude that $\la \delta s_i(t) \ra = 0$, and hence in the first order $\la \delta C_i \ra = 0$. Consequently
\begin{align}
    \label{eq:stochastic_trigger_price}
    C_{st} = \la C \ra = C_0 + \la \delta C \ra \approx C_0\,,
\end{align}
which means that the stochastic trigger price equals the intrinsic trigger price. Since any time moment $t$ during the exercise period can be considered as a starting point, the following statement holds: for any time $t$ the optimal stochastic exercise is bang-bang, and the stochastic trigger price $C_{st}(t)$ is approximately equal to the intrinsic trigger price $C_0(t)$, calculated at time $t$ on the basis of the deterministic forward curve $F(t,T)$.

Note that to derive the relation (\ref{eq:stochastic_trigger_price_1}) we did not use any approximation, and hence this relation is exact. For derivation of relation~(\ref{eq:stochastic_trigger_price}) we used the only assumption of small variation
\begin{align}
    \label{eq:condition_for_rolling intrinsic}
    \frac{ \sqrt{\lla \delta s^2(T_e) \rra}}{s_0(T_e)}\ll 1\,.
\end{align}
For a 1-factor mean reversion price process the relative spot price variation can be easily estimated:
\[\frac{ \sqrt{\lla \delta s^2(T_e) \rra}}{s_0(T_e)} =
 \sqrt{ \exp \left[ \frac{\sigma_0^2}{2\,\alpha}\,\Big(1- e^{-2\,\alpha\,T_e}\Big)\right]-1 }\,.\]
In particular, for $\alpha\,T_e\ll 1$ the condition for the intrinsic exercise approximation becomes
\[\sqrt{ \exp(\sigma_0^2\,T_e) -1} \ll 1\,,\]
and for $\alpha\,T_e\gg 1$ we get the condition
\[ \sqrt{ \exp \left(\frac{\sigma_0^2}{2\,\alpha}\right) - 1} \ll 1\,. \]

\subsection{Variation of the intrinsic target function}

In this section we find a variation of the intrinsic target function on the constraint surface. We take into account the dependency between variations of the forward curve and the intrinsic trigger price, as in Eq.~(\ref{eq:dC_2_order}).

Below we omit the argument $t$ and use the notation $F(T)\equiv F(t,T)$. We write the intrinsic target function in the form
\begin{align}
    S(t) = -\int_t^{T_e} r(T)\,F(T) \, dT\,;\qquad \text{where} \quad
    r(T) = r_{min} + \Delta r\,\theta(C - F(T))\,.
\end{align}
Next we use the expansion
\begin{align}
    \delta S = \delta_C S + \delta_F S + \frac12( \delta_{FF} S + 2\,\delta_{CF} S + \delta_{CC} S)\,.
\end{align}
The variational derivatives can be found easily:
\begin{align}
    & \delta_C S = \frac{\p S}{\p C}\,\delta C = -C\,K\,\,\delta C\,; \\
    & \delta_F S = \int \frac{\delta S}{\delta F(T)}\,\delta F(T)\,dT =
        \int \Big( C\,r^{(1)} - r \Big)\,\delta F\,dT\,;
\end{align}
\begin{align}
    & \delta_{FF} S = \iint \frac{\delta^2 S}{\delta F(u)\,
        \delta F(v)}\,\delta F(u)\,\delta F(v)\,du\,dv =
         \int \Big( 2\,r^{(1)} - r^{(2)}\,F \Big)\,L_{uu}\,du \,; \\ 
    & \delta_{FC} S = \int \frac{\delta^2 S}{\delta F(T)\,\p C}\,\delta F(T)\,dT\,\,\delta C =
    \frac{1}{K} \iint \Big( r^{(2)}_u\,F_u - r^{(1)}_u \Big)\,r_v \,
        L_{uv}\,\,du\,dv = \nonumber \\
    & \hspace{6cm} =\frac{C}{K} \iint r^{(2)}_u\,r^{(1)}_v\,L_{uv}\,\,du\,dv \,;\\
    & \delta_{CC} S = -\int r^{(2)}\,F \,dt\,\,\delta C^2 = - \frac{K + C\,M}{K^2}
        \iint r^{(1)}_u\,r^{(1)}_v\,L_{uv}\,\,du\,dv\,;
\end{align}
where $L_{uv} = \delta F(u)\,\delta F(v)$, $L_{uu} = \delta F^2(u)$.

Combining all variation terms we finally obtain:
\begin{align}
    \label{eq:deltaS}
    \delta S = & -\int r(T) \,\delta F(T)\,dT + \frac12 \int r^{(1)}\, L_{uu} \,du -
         \frac{1}{2\,K} \iint r^{(1)}_u\,r^{(1)}_v \,L_{uv}\, \,du\,dv \,.
\end{align}
We can easily interpret the first term appearing in the latter expression. The variation
\[
    -\int r(T)\,\delta F(T)\,dT
\]
reflects the change of the value of the current hedge volume of underlying due to the changed forward prices, provided the hedge volume equals the optimal intrinsic exercise $r(T)$.

Next non-vanishing term is of the second order:
\[
    \frac12 \int r^{(1)}(T)\,\delta F^2(T)\,dT =
        \frac{\Delta r}{2} \int \delta(C-F(T))\,\delta F^2(T)\,dT\,.
\]
Due to the delta-function under the integral this term is reduced to a sum over trigger times $\{T_i: F(T_i) = C \}$. This term appears due to the change of the perturbed trajectory $r(T)+\delta r(T)$. It can be shown (see Sec.~(\ref{sec:delta_r})) that the perturbed trajectory almost everywhere coincides with the unperturbed one $r(T)$ except a few points, where the transition between injection and release takes place. These points are the trigger times. Change of the strategy on an infinitely small period of time around every trigger time leads to the second order correction of the target function.

It is worth noticing that the second order correction is always non-negative. This reflects the fact that the rolling intrinsic strategy leads to a non-negative drift of the storage value (see below).

The term
\[
     \frac{1}{2\,K} \iint r^{(1)}_u\,r^{(1)}_v \,L_{uv}\, \,du\,dv
\]
takes into account the cross-correlations between the forward price returns for different maturities.

\subsubsection{Interpretation in financial terms}

In the financial calculus the derivatives of the target function with respect to the different parameters deserve a special attention. The most relevant of them for the portfolio management are usually designated by Greek letters, and traditionally called all together ``Greeks''.

One of the most important derivatives is the derivative of the option value $V$ with respect to the underlying price $F$:
\[\Delta = \frac{\p V}{\p F}\]
It is termed ``delta'' and can be used for constructing a delta-neutral portfolio -- such a portfolio, which value does not change under infinitesimal changes of the price.

The second derivative of the option value with respect to the spot price is termed ``gamma'':
\[\Gamma = \frac{\p^2 V}{\p F^2} \,.\]
The gamma is a measure of the curvature of the target function with respect to the underlying price. It is also responsible for the time value of the option.

Combining first and second derivatives, the variation of the option value can be represented up to the second order as
\begin{align}
     dV = \Delta\,dF + \frac12\,\Gamma\,dF^2\,.
\end{align}

Comparing the latter equation with Eq.~(\ref{eq:deltaS}) we conclude that the first order term in the expansion represents the option ``delta'' (within the intrinsic strategy approximation):
\begin{align}
    \Delta \sim - \int  r(u)\,\delta F(u) \,du\,,
\end{align}
and the second order term represents the storage option ``gamma'':
\begin{align}
    \Gamma \sim \int  r^{(1)}(u)\,L_{uu}\,du -
    \frac{1}{K} \iint r^{(1)}(u)\,r^{(1)}(v) \, L_{uv} \,\,du\,dv\,.
\end{align}

\subsection{Variation of the intrinsic extremal trajectory}
\label{sec:delta_r}

Here we find a variation of the optimal intrinsic trajectory $r(T)$ on the constraint surface. We remind that the optimal intrinsic trajectory can be represented as
\[
    r(T) = r_{min} + \Delta r\,\theta(C - F(T))\,,
\]
where we have omitted the observation time $t$. Expanding to the second order in $F$ and $C$ we obtain:
\begin{align}
    \delta r(T) = \frac{\p r(T)}{\p F}\,\delta F(T)
    + \frac{\p r(T)}{\p C}\,dC
    + \frac12\,\frac{\p^2 r(T)}{\p F^2}\,\delta F^2(T)
    + \frac{\p^2 r(T)}{\p F\, \p C}\,\delta F\, dC
    + \frac12\,\frac{\p^2 r(T)}{\p C^2}\, dC^2 \, .
\end{align}
Evaluating the partial derivatives and making use of Eq.~(\ref{eq:dC_2_order}) we finally get
\begin{align}
    \label{eq:delta_r}
    \delta r(T)& =  -r^{(1)}\,\delta F + \frac12\,r^{(2)}\,L_{tt}  -
    \frac{r^{(2)}}{K}\,\delta F \int r^{(1)}_v\,\delta F_v   +
    \frac{r^{(2)}}{2\,K^2}\iint r^{(1)}_u\,r^{(1)}_v\,L_{uv}  +\nonumber \\
    &      \frac{r^{(1)}}{K}\,\Big(
            \int r^{(1)}_u\,\delta F_u
            - \frac12 \int r^{(2)}_u\,L_{uu}
            + \frac{1}{K}\iint r^{(2)}_u\,r^{(1)}_v\,L_{uv}
            - \frac{M}{2\,K^2}\,\iint r^{(1)}_u\,r^{(1)}_v\,L_{uv}
        \Big).
\end{align}

We see that $\delta r(T)$ has singularities at the trigger times $\{T_i:\ F(T_i) = C\}$. We need to give an interpretation to this fact, since the trajectory correction $\delta r(T)$ in this form violates the maximum injection/release constraints. We can show that the singular trajectory variation can be interpreted as a shift of trigger times $T_i$ by some value $\delta T$.

We can give an interpretation to the obtained singularity of $\delta r$, if we consider it under an integral $\int \delta r(T) \,\phi(T)\,dt$, where $\phi(T)$ is an arbitrary function of time. Approximately  the $\delta-$function can be replaced by a step function with support $\delta t$ and height $\Delta r$. The duration $\delta T$ should be found to fit the integral value.

Let us consider $\delta r$ in a form
\[\delta r(T) = k\,r^{(1)}(T) = k\, \Delta r\,\delta(C-F(T))\,. \]
Let $T^*$ be a trigger time. Then the integral around this time gives:
\[
    \int_{T^*-\epsilon}^{T^*+\epsilon} \delta r(T) \, \phi(T)\,dT =
    \frac{k\,\Delta r}{|\dot F(T^*)|}\,\phi(T^*)\,.
\]
As stated above, the variation $\delta r(T)$ around the trigger time $T^*$ can be approximately replaced with a step function $\eta(T)$ with support $\delta T$ and height $\Delta r$. The width $\delta T$ can be estimated from simple consideration:
\[
    \int_{T^*-\epsilon}^{T^*+\epsilon} \delta r(T) \,\phi(T)\,dT =
    k\,\Delta r\,\frac{\phi(T^*)}{|\dot F(T^*)|} \approx
    \Delta r\,\delta T\,\phi(T^*) \,.
\]
We thus find for the absolute value of time shift:
\begin{align}
    \delta T \approx \left| \frac{k}{ \dot F(T^*)}\right|\,.
\end{align}
Since the time $t^*$ just splits the regions of $\dot q = r_{min}$ and $\dot q = r_{max}$, the previous consideration has a meaning, that the singularity of $\delta r$ can be interpreted as shifting of the trigger time $T^*$ by the value~$\delta T$ to the right or left. If $k > 0$, then the trigger time is always shifted in the direction of injection. If $k < 0$, then the trigger time is always shifted in the direction of release. We can represent the time shift, which takes the sign into account:
\begin{align}
    \label{eq:time_shift}
    \delta T = - \frac{k}{ \dot F(T^*)}\,.
\end{align}

An important property of $\delta r$ is that the integral $\int \delta r(T)\,dT$ must vanish. Indeed, from Eq.~(\ref{eq:constraint}) we conclude that
\[
    \delta Q_{end} = \int \delta r(T)\,dT = 0\,.
\]
Integrating Eq.~(\ref{eq:delta_r}) and taking into account the definitions $K = \int r^{(1)}(T)\,dT$ \\ and  $M = \int r^{(2)}(T)\,dT$ we find
\begin{align}
  & \int \delta r(T)\,dT = 0\,.
\end{align}
This proves the consistency of the expansion formula (\ref{eq:delta_r}).

\subsection{Dynamic hedge and time value}

As a ``delta'' of a vanilla option we understand a derivative of the option value with respect to the price of the underlying. This derivative has a dimension of volume and has an obvious meaning: this is an amount of the underlying, whose value changes (up to the first order) to the same extent under the small variation of the price, as the option value itself.

The delta can be used for the delta-hedging. If the holder of the option takes a forward (hedge) position, which equals the delta in its amount, but has an opposite sign, then the combination of both -- option and hedge position -- has vanishing derivative with respect to the price. This means that for a very small variation of the price, i.e. for a very short period of time, the value of the portfolio remains constant (up to the first order). In this case we say that the portfolio is delta-neutral. Since the delta itself depends on the price, the position, which was delta-neutral at some time point, will not be so after a short period of time. To remain delta-neutral, one has to recalculate the delta and update the hedge position continuously. In an ideal situation, if the update of the hedge position can be done continuously, and if the market friction can be neglected, the value of the portfolio $\Pi$ (which includes the option value, the value of the hedge volume and cumulative cash flows) remains constant.

The similar logic applies to the storage option. We can identify such a volume of underlying, which we can use as a hedge position. But unlike the vanilla option the storage option has an array of maturity times, which can be thought of as an array of different hedge products, each having a different forward price $F_k = F(T_k)$. For every delivery time $T_k$ we can calculate a derivative $\p V(F)/\p F_k$ with respect to the forward price on that particular maturity. The delta in this case becomes an array:
\begin{align}
    \Delta_k = \frac{\p V(F)}{\p F_k}\,,
\end{align}
which is a discrete analogue of a functional derivative $\Delta(T) = \delta V/\delta F(T)$.

At the beginning of the storage contract the portfolio consists only of the storage option. The initial portfolio value (sometimes referred to as ``Profit and Loss'' - P\&L) equals the estimated option value. As the storage starts operating, the owner of the storage sells and buys the underlying and changes the hedge position. The P\&L changes according to the actual cash-flows, option value and the value of the hedge position. At the end of the storage contract the terminal P\&L consists only of the cumulative cash-flow, since no open hedge position is left after the end of the storage period. If the hedge position is optimal at any time, then the P\&L remains constant, and the terminal P\&L equals the initial one, i.e. the storage owner earns exactly the amount predicted by the stochastic storage model.

If the storage delta can not be calculated exactly, then the P\&L becomes volatile, fluctuating around some expected value. Depending on the ``quality'' of the hedge position, the deviation of the terminal P\&L from the expected value may become smaller or bigger.

Below we show that the expected value of the terminal P\&L depends exceptionally on the prompt exercise trades (i.e. on the value of the delta at the current time $T=t$), whereas the width of the distribution of the terminal P\&L depends on the ``quality'' of the hedge position. Whatever the hedge strategy is used, the correct average time value is obtained only if at every time moment the prompt exercise $\dot q(t)$ is optimal.

In the next section we consider the rolling intrinsic exercise strategy as one of the possible hedge strategies. The rolling intrinsic strategy has a number of remarkable properties. One important feature of the rolling intrinsic is that it provides an (almost everywhere) optimal exercise decisions, and hence guarantees that the expectation value of the P\&L equals the true option value. Rolling intrinsic also allows to estimate the option time value by calculating the cumulative cash flows from the hedge and exercise trades.

\subsubsection{Rolling Intrinsic strategy}

The deterministic consideration of the storage problem -- the intrinsic value -- gives a good first order estimate of the storage value. The intrinsic value can be computed for example by a dynamic programming algorithm, and has a remarkable property of the trigger price: the optimal intrinsic exercise strategy is bang-bang, each bang triggered by crossing the trigger level by the forward price curve.

The stochastic reality makes the true value of the storage option different, adding to it a time value. It appears due to additional profit opportunities on the price fluctuations. As discussed in the Sec.~\ref{sec:4.6.4}, under some assumptions a good approximation for the ``stochastic'' trigger price is the intrinsic one. It means that an optimal prompt exercise decision at any time moment $t$ can be made by means of trigger level $C_{st}(t)$ for which the intrinsic level $C_0(t)$ is a good approximation.

In Sec.~(\ref{sec:stoch_opt_problem}) we have introduced the profit function $P$, which has a meaning of the cumulative cash flow from all hedge and exercise trades, and formulated the stochastic optimisation problem in terms of probabilistic maximisation of the terminal profit. We repeat the same argumentation here in more details in application to the rolling intrinsic strategy.

Let $F(t,T)$ be the forward price observed in the market at time $t$. Let also $r(t,T)= \dot q(t,T)$ be the intrinsic optimal exercise trajectory calculated on the forward price curve $F(t,T)$. The volume of future trading $r(0,T)$ can be interpreted as the intrinsic hedge volume. Prior the exercise period the whole volume $r(0,T)$ can be traded forward, providing the owner of the storage option the guaranteed profit
\[
    P(0) = -\int_0^{T_e} r(0,T)\,F(0,T)\,dT\,,
\]
which equals the intrinsic value. Although strictly speaking the cash-flow from the forward contracts takes place at maturity times, we simplify the picture by assigning all the future cash-flows to the time moment the forward (hedge) position is taken. Thus, we can say that the intrinsic value is locked at the very first time moment.

During the exercise period the breathing prices cause the change of optimal intrinsic strategy. If at time $t$ the price curve is  $F(t,T)$, then at time $t+dt$ it becomes different
\[
    F(t+dt,T) = F(t,T) + \delta F(t,T)\,,
\]
leading to variation of optimal intrinsic strategy
\[
    r(t+dt,T) = r(t,T) + \delta r(t,T)\,.
\]
The volume $\delta r(t,T)$ can be traded continuously (dynamic hedge), leading to continuous change of the profit function:
\begin{align}
    \label{eq:delta_S}
    \delta P(t) = -\int_t^{T_e} \delta r(t,T) \Big(F(t,T) + \delta F(t,T) \Big)\,dT
\end{align}
corresponding to time period $dt$ (again we assign the cash-flows from all future trades to the current time moment $t$). The described strategy is called ``rolling intrinsic''. It assumes that at every time moment the option holder takes the hedge position equal to the current optimal intrinsic strategy.

Since the trajectory $r+\delta r$ is optimal for the price curve $F+\delta F$, the following inequality must hold (we omit the limits of integration and observation time $t$)
\begin{align}
     & - \int \Big(r(T) + \delta r(T) \Big) \Big(F(T) + \delta F(T) \Big) \,dT \geq
     - \int r(T) \, \Big(F(T) + \delta F(T) \Big) \,dT\,,
\end{align}
from which follows
\begin{align}
    \label{eq:PnL_inequality}
     \delta P(t) \geq 0\,.
\end{align}
Consequently the dynamic hedge in the rolling intrinsic strategy can only increase the value of the profit function. In other words, the time value of the storage option is positive (non-negative).

Interestingly, the inequality (\ref{eq:PnL_inequality}) is only valid for rolling intrinsic strategy.

In the appendix~\ref{ap:A2} we show how the time value of the storage option can be calculated from the cumulative cash flows of a rolling intrinsic hedge and exercise trades. In the following sections we apply this technique to estimate the time value of the storage option.

\subsubsection{Meaning of the Dynamic Hedge}

In this section we show that the hedge position is not relevant for the expected value of the storage option. The only prompt exercise trade, i.e. the spot trade, is crucial for the expected value of the terminal P\&L. The hedge position only effects the distribution of the terminal profit.

Let us consider some delivery time $T\in[0,T_e]$. Let also at some observation time $t<T$ the optimal hedge position be
\[h_{opt}(t,T)\,.\]
We do not specify here how to calculate the optimal hedge position $h_{opt}(t,T)$. We only believe that it could be calculated in some way. Suppose that the real hedge position, which is taken at time $t$
\[h(t,T)\]
is not necessarily equal the optimal one:
\[h(t,T)\not = h_{opt}(t,T)\]
almost everywhere. On the other hand, we demand that at the beginning the hedge position is zero, and as the time approach the maturity, the hedge position has to be equal the optimal one:
\[h(0,T) \equiv 0\,;\qquad h(T,T) \equiv h_{opt}(T,T) \qquad \text{for all } T \,.\]
The volume of the underlying associated with the hedge position for the delivery period $[T, T+\delta T]$ is
\[ h(t,T)\,\delta T\,. \]
Since the initial hedge position is everywhere zero, there is no initial cash-flow, associated with the hedge position. The hedge position changes with the time $t$. The increment of the hedge position
\[dh(t,T) = \frac{\p}{\p t}h(t,T)\,dt\]
leads to the cash-flow
\[ dP(t,T) = - \Big(F(t,T)+dF(t,T)\Big)\,dh(t,T)\,\delta T\,,\]
where $F(t,T)$ is the forward price for the delivery on $T$, observed on $t$. The total cash-flow per delivery period $\delta T$ becomes
\begin{align}
    \label{eq:expected_cashflow}
    \frac{P(T)}{\delta T} = -\int_0^T F(t,T)\,dh(t,T)-\int_0^T  dF(t,T)\,dh(t,T) \,.
\end{align}


Now we apply integration by parts, which In Ito calculus has to be slightly modified. Indeed, let
\[ g(t) = y(t)\,z(t)\,,\]
where $z(t)$ and $y(t)$ are correlated stochastic processes. We use the Ito formula to obtain
\begin{align}
    \label{eq:ItoChainRule}
    dg = \frac{\p g}{\p y}\,dy + \frac{\p g}{\p z}\,dz
        + \frac12\,\frac{\p^2 g}{\p y^2}\, dy^2 
        + \frac{\p^2 g}{\p y\,\p z}\,  dy\,dz 
        + \frac12\,\frac{\p^2 g}{\p z^2}\, dz^2  = z\,dy + y\,dz +  dy\,dz \,.
\end{align}
Integrating from 1 to 2 we obtain the modified ``integration by parts'' formula:
\begin{align}
    \left. (y\,z) \right|_1^2  = \int_1^2 y\,dz + \int_1^2 z\,dy + \int_1^2  dy\,dz\,;
\end{align}
Applying this formula to Eq.~(\ref{eq:expected_cashflow}) we obtain
\begin{align}
    \frac{1}{\delta T} P(T) =
    -  \left. \Big( F(t,T)\,h(t,T)\Big) \right|_{t=0}^{t = T}  +
    \int_0^T  h(t,T) \, dF(t,T) \,.
\end{align}
Averaging the obtained expression over the price increments $dF$, and noticing that
\[\la h(t,T) \, dF(t,T) \ra =  h(t,T) \, \la dF(t,T) \ra = 0\,,\]
the expected cash-flow becomes
\begin{align}
    \frac{1}{\delta T}\la P(T)\ra &= - \la F(T,T)\,h(T,T)\ra= - \la F(T,T)\,h_{opt}(T,T)\ra\,.
\end{align}
Thus, the expected cash-flow is independent on the hedge strategy $h(t,T)$ for $t<T$, and only depends on the prompt exercise trade $h(T,T)$.

From this an important consequence follows: the rolling intrinsic strategy is a good approximation of the exact optimal strategy. Indeed, as has been shown in Sec.~\ref{sec:4.6.4} the intrinsic prompt exercise is a very good approximation of the stochastic one, and hence, rolling intrinsic leads to a very good expected value of the terminal storage profit. 

On the other hand, the intrinsic hedge profile may be not close to the optimal hedge profile, and thus it may lead to rather poor efficiency of hedge in terms of stabilising the profit function. We can estimate the ``quality'' of the intrinsic hedge strategy by calculating the standard deviation of the terminal value of the profit function.

\subsection{Time Value}

\newcommand{\dF}{ {\delta F} }

For the rolling intrinsic strategy the increment of the target function is given by Eq.~(\ref{eq:delta_S}). Averaging the increment $\delta P(t)$ and its square with respect to the price increment $\dF$ we obtain:
\begin{align}
    \label{eq:4.55}
    &\la \delta P \ra_\dF = - \int_t^{T_e} \la \delta r(T) \ra_{\dF}\, F(T)\,dT -
        \int_t^{T_e} \la \delta r(T)\,\delta F(T) \ra_{\dF}\,dT \,; \\
    \label{eq:4.56}
    & \la \delta P^2 \ra_{\dF} =
        \iint_t^{T_e} \la\delta r(T')\,\delta r(T'')\ra_{\dF}\,F(T')\,F(T'')\,dT'\,dT''\,,
\end{align}
where $\la \cdot \ra_{dF}$ is the averaging over the stochastic increment $dF$. 

Both $\la \delta P \ra_{\delta F}$ and $\lla \delta P^2 \rra_{\delta F}$ are proportional to $\delta t$. Designating
\begin{align}
    \mu_p(t) = \frac{\la \delta P(t) \ra_{\dF}}{\delta t}\,;\qquad
    \sigma_p^2(t) = \frac{\la \delta P^2(t) \ra_{\dF}}{\delta t}\,,
\end{align}
we can assign to $P(t)$ an effective stochastic process
\begin{align}
    dP(t) = \mu_p(t)\, dt + \sigma_p(t) \,dW_t\,,
\end{align}
Notice that the average $\la P \ra_{\delta F}$ as well as $\mu_p$ and $\sigma_p$ are stochastic values, since they depend (implicitly or explicitly) on the stochastic price $F(T)$. The drift $\mu_p$ and volatility $\sigma_p$ averaged over the forward price will allow us to find the distribution parameters of the terminal profit~$P(T_e)$. Indeed, the expectation value and the variance of the time value $V_T = P(T_e) - P_0$ are given by (see App.~\ref{ap:A2})
\begin{align}
     &\la V_T \ra = \int_0^{T_e} \bar \mu_p(t)\,dt  \\
     &\lla V_T^2 \rra = \int_0^{T_e} \bar \sigma^2_p(t)\,dt
\end{align}
where we have designated
\[
    \bar\mu_p = \la \mu_p \ra_F\,;\qquad \bar\sigma_p^2 = \la \sigma_p^2 \ra_F\,.
\]

Substituting $\delta r(T)$ into Eq.~(\ref{eq:4.55}) and performing integration we obtain
\begin{align}
    \label{eq:delta_P}
    &\la \delta P \ra_\dF = \frac12 \int r^{(1)}(u)\,L_{uu} \,du -
        \frac{1}{2\,K} \iint  r^{(1)}_u\,r^{(1)}_v\,L_{uv}\,\,du\,dv  \,;
\end{align}
where the integration is performed from $t$ to $T_e$. This result could also be obtained from Eq.~(\ref{eq:deltaS}), since (see App.~\ref{ap:A2})
\begin{align}
    \label{eq:dP=dS}
     \la \delta P \ra_\dF = \la \delta S \ra_\dF\,.
\end{align}

Next we observe that $r^{(1)}(T) = \Delta r\,\delta(C-F(T))$ as well as $r^{(2)}(T) = \Delta r\,\delta'(C-F(T))$ are singular functions, and the integrals of the type $\int r^{(1)}(T)\,f(T)\,dT$ and $\int r^{(2)}(T)\,f(T)\,dT$ are reduced to a summation over the set of trigger times $\{T_i: F(T_i) = C\}$.  Making use of the identities (\ref{eq:A4}) and (\ref{eq:A5}) we obtain:
\begin{align}
    \label{eq:mu}
    & \mu_p(t) = \frac{\Delta r}{2\,\delta t} \sum_i \frac{L_i(t)}{|\dot F_i(t)|} -
    \frac{\Delta r^2}{2\,K(t)\,\delta t}\sum_{ij} \frac{L_{ij}(t)}{|\dot F_i(t)|\,|\dot F_j(t)|}\,,
\end{align}
where
\begin{align*}
    & L_{ij}(t) = L(t,T_i,T_j)\,;\quad L_i(t) \equiv L_{ii}(t)\,; \\
    & \dot F_i(t) = \left. \frac{\p}{\p T} F(t,T)\right|_{T = T_i} \,;\\
    & K(t) = \Delta r \sum_{k>t} \frac{1}{|\dot F_k(t)|}\,.
\end{align*}
This value is stochastic for $t>0$. Using the results of App.~\ref{app:averaged_integral} and using the notation
\begin{align}
    & \Lambda_{ij}(t) = \frac{1}{\delta t}\,\la L_{ij}\ra\,; \qquad
    \Lambda_i(t) := \Lambda_{ii}(t)\,;
\end{align}
we finally obtain the average of $\mu_p$ with respect to the forward curve $F$:
\begin{align}
    \label{eq:av_mu_storage}
    & \bar \mu_p(t) \approx \frac{\Delta r}{2} \sum_i \frac{ \Lambda_i(t)}{|\dot F_{0i}(t)|} -
    \frac{\Delta r^2}{2\, K_0(t)}\sum_{ij} \frac{\Lambda_{ij}(t)}{|\dot F_{0i}(t)|\,|\dot F_{0j}(t)|}\,; \quad
    \text{where}\quad K_0(t) = \Delta r \sum_{k>t} \frac{1}{|\dot F_{0k}(t)|}\,.
\end{align}

Using the results from the App.~\ref{app:averaged_integral} the latter summation formula can under some assumptions be replaced with the integration, which would allow an easier analytic approximation.

\subsection{Example calculation}
\label{sec:example_calculation_storage}

We apply now the expansion formula to a simple toy example of a storage contract with one factor price process (with $\beta= 0$). The correlation function for this price process is given by
\[
    \Lambda(t,T_1,T_2) = \sigma_0^2\,e^{-\alpha(T_1-t)}\,e^{-\alpha(T_2-t)}\,\la F(t,T_1)\,F(t,T_2)\ra 
    \approx\sigma_0^2\,e^{-\alpha(T_1-t)}\,e^{-\alpha(T_2-t)}\,F_0(T_1)\,F_0(T_2) \,.
\]
As initial price condition we consider a periodic function
\[
    F_0(T) = F_c + \Delta F\,\sin(\omega \,T)\,;\qquad \omega = \frac{\pi\,N}{T_e} = \frac{\pi}{\Delta T}\,,
\]
where $T_e$ is the exercise period, $N$ is the number of trigger times and $\Delta T$ is the distance between the nearest trigger times.
Let also
\[C = F_c\]
be the initial trigger level. We obtain:
\begin{align}
    T_i = \frac{i\,T_e}{N}\,;\qquad
    F_{0i} = F_c \,;\qquad
    |\dot F_{0i} | = \Delta F\,\omega = \pi\,\frac{\Delta F}{\Delta T}\,; \qquad
    \ddot F_i = 0\,.
\end{align}

Next we substitute
\begin{align*}
    & \Lambda_{ij} = \sigma_i\,\sigma_j\,F_{0i}\,F_{0j} = F_c^2\,\sigma_i\,\sigma_j\,;\qquad \text{where}\quad
    \sigma_i = \sigma(t,T_i) = \sigma_0\,e^{-\alpha(T_i-t)}\,; \\
    & \Lambda_i = F_c^2\,\sigma_i^2\,; \\
    & K(t) = \frac{\Delta r}{\pi\,\Delta F}\sum_i \Delta T =
        \frac{\Delta r}{\pi\,\Delta F}\,(T_e-t)
\end{align*}
into Eq.~(\ref{eq:av_mu_storage}) to obtain
\begin{align}
  \bar\mu_p(t) =& \frac{\Delta r\,F_c^2}{2\pi\,\Delta F}\,\sum_i \sigma_i^2\,\Delta T -
  \frac{\Delta r^2\,F_c^2}{2\,\pi^2\,\Delta F^2\,K(t)}\,\sum_i \sigma_i\,\Delta T\,\sum_j \sigma_j\,\Delta T
  \approx \nonumber  \\
  & \hspace{1cm} \approx \frac{\Delta r\,F_c^2\,\sigma_0^2}{2\,\pi\,\Delta F}
  \left[\int_0^{T_e-t} e^{-2\,\alpha\,T}\,dT -
  \frac{1}{T-t}  \left( \int_0^{T_e-t} e^{-\alpha\,T}\,dT \right)^2 \right]\,.
\end{align}
Performing integration we get:
\begin{align}
    \label{eq:mu_final}
    \bar \mu_p(t) =&
    \frac{\Delta r\,F_c^2\,\sigma_0^2}{2\,\pi\,\Delta F}\,
    \left[ \frac{ 1 - e^{-2\,\alpha\,(T_e-t)} }{2\,\alpha} -
        \frac{1 - 2\,e^{-\alpha\,(T_e-t)} +  e^{-2\,\alpha\,(T_e-t)} }{\alpha^2\,(T_e-t)}\,
    \right]\,.
\end{align}

For the option time value we finally obtain
\begin{align}
    \label{eq:storage_time_value}
    & V_T := \la P(T_e)\ra  - P_0  = \int _0^{T_e} \bar \mu_p(t) \,dt =
             \frac{\Delta r\,F_c^2\,\sigma_0^2\,T_e^2}{8\,\pi\,\Delta F} \, \Phi(\alpha\,T_e)\,,
\end{align}
where we have introduced a function
\begin{align}
     \Phi(x) = \frac{1}{x^2}\,\left(
            e^{-2\,x} - 1 + 2\,x -4\,\gamma - 8\,\Gamma(0,x) + 4\,\Gamma(0,2\,x) + 4\,\ln\frac{2}{x}
        \right)\,.
\end{align}
Here $\gamma\approx 0.577$ is the Euler's constant
and $\Gamma(a,z)$ is incomplete gamma function
\[\Gamma(a,z) = \int_z^\infty t^{a-1}\,e^{-t}\,dt\,.\]

\subsubsection{Time Value as a function of $\alpha$}

\begin{figure}[htb!]
  \begin{center}
    \includegraphics[angle= 0, width=0.9\columnwidth] {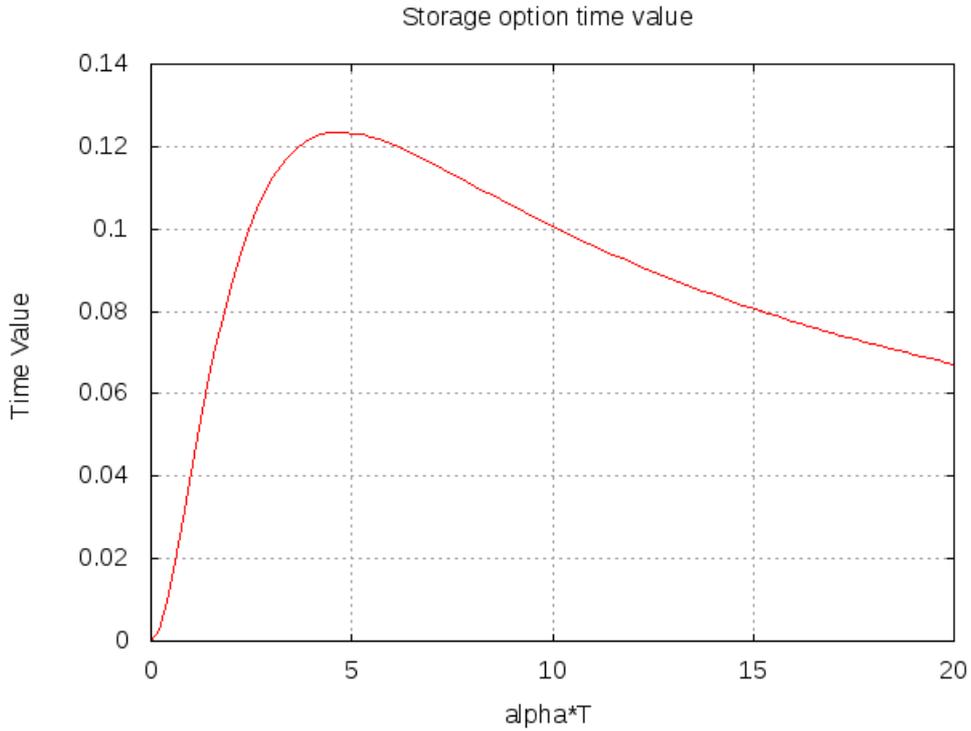}  
    \caption{Time Value $\Phi$ as a function of $\alpha\,T_e$}
    \label{fig:2}
  \end{center}
\end{figure}

The function $\Phi(x)$ is bell-shaped with $x^2$ asymptotics on the left wing as $x\to +0$ and $1/x$ asymptotics on the right wing as $x\to\infty$ (see Fig.~\ref{fig:2}). The function reaches it's maximum approximately at $x^*\approx 5.04$ and reaches there the value
\[\Phi(x^*) \approx 0.12 \]
Thus, if we consider the time value $P_t$ as a function of the parameter $\alpha$ (the time $T_e$ is kept constant), then it reaches its maximum at
\begin{align}
    \alpha^* \approx \frac{5}{T_e}\,,
\end{align}
and the maximum time value is
\begin{align}
    V_T^* \approx \frac{\Delta r\,F_c^2\,\sigma_0^2}{67\,\pi\,\Delta F}\,T_e^2\,.
\end{align}

The function $\Phi(x)$ in the whole range can approximately be represented in the form
\begin{align}
    \Phi(x) \approx k(x)\,\frac{-5 + e^{-2\,x} + 2\,x + 4\,e^{-x}\,(1+x)}{x^2}\,,
\end{align}
where $k(x)$ could roughly be considered as constant. More precisely, $k\approx 0.5$ to better fit the left wing of $\Phi$ and $k\approx 0.9$ to better fit the right wing. Around maximum the best fit is achieved with $k\approx 0.6$. In the whole range we can take $k\approx 0.9-0.4\,e^{-x/18}$.

This approximation results from the replacement
\[
 \frac{1-2\,e^{-x} + e^{-2\,x} }{x} \approx x\,e^{-x}\,.
\]
in Eq.~(\ref{eq:mu_final}).

\subsubsection{Limiting case $\alpha\,T_e \ll 1$}

For small $x$ the function $\Phi(x)$ has an asymptotics
\[\Phi(x) \approx \frac{x^2}{12}\,,\qquad x\to 0\,.\]
Thus, for small $\alpha$ satisfying the inequality $\alpha \ll \frac{1}{T_e}$ the asymptotic time value becomes
\begin{align}
    V_T \approx \frac{\Delta r\,F_c^2\,\sigma_0^2}{96\,\pi\,\Delta F}\,\alpha^2\,T_e^4\qquad
    \alpha\,T_e \to 0\,.
\end{align}
We conclude that the storage option time value for $\alpha\to 0$ is proportional to $\alpha^2$. In particular, the time value vanishes for $\alpha=0$.

\subsubsection{Limiting case $\alpha\,T_e \gg 1$}

For this asymptotics we obtain
\begin{align}
    \Phi(x) \approx \frac{-0.5 + 2\,x - 4\,\ln x}{x^2} \approx \frac{2}{x}\,.
\end{align}
Thus, for big $\alpha$ satisfying the inequality $\alpha \gg \frac{1}{T_e}$ the asymptotic time value becomes
\begin{align}
    V_T \approx \frac{\Delta r\,F_c^2\,\sigma_0^2}{4\,\pi\,\Delta F}\,\frac{T_e}{\alpha}\,,\qquad
    \alpha\,T_e \to \infty\,.
\end{align}

\subsubsection{Comparison of analytical results with numerical simulation}

\begin{figure}[htb!]
  \begin{center}
    \includegraphics[angle= 0, width = 0.8\textwidth] {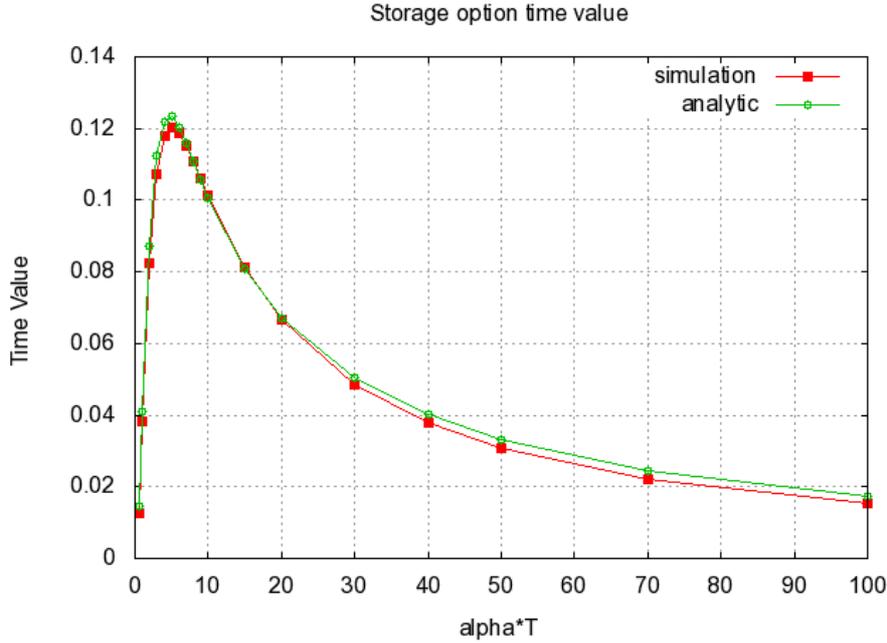}  
    \caption{Time Value $\Phi$ as a function of $\alpha\,T_e$. Analytic solution is compared with numerical simulation.}
    \label{fig:3}
  \end{center}
\end{figure}

The comparison of the analytical formula for the time value with numerical simulation can be seen in the Fig.~\ref{fig:3}.

For this graph we have created a toy storage contract with deliberately reduced influence of the surface effects. To a large extent the difference can be explained by the time discretisation effect. The analytical formula is obtained in the continuous time approximation, whereas in the real life the time is discrete (for instance, gas delivery is usually nominated once a day, and hence the smallest time step is one day). The discrete time system has less trading opportunities than the continuous time system, and hence its exercise is ``suboptimal'' from the continuous time perspective. For this reason the discrete time system has slightly smaller time value than the continuous one.

Other sources of error are surface effects (boundary influence), higher order corrections of the exercise strategy and some approximations used to simplify the analytic expression.

\subsection{Swing option}
\label{sec:swing_option}

In the previous sections we have considered the storage option problem with fixed terminal condition. This condition has a significant impact on the option value. Indeed, this is simply an additional constraint which restricts the possible strategies and hence can only decrease the option value. Although in most of the storage contracts there is no restriction on the possible terminal level, the problem still can be considered as having the fixed terminal state constraint. The reason for that is simple. The derivative of the storage value with respect to the terminal state is given by Eq.~(\ref{eq:spacial_derivative})
\[ \frac{\p S}{\p q_{end}} = -C \]
where $C$ is the trigger price of the last part of the trajectory. If the prices $F(T)$ are strictly positive, the trigger price can only be positive as well, since the forward curve should cross the trigger level at some point\footnote{Strictly speaking, we can construct some extreme example, when the trigger price can be zero. For instance if the storage at the beginning is fool and release costs are so high that for some periods of time it is profitable rather to do nothing than to release gas. Here we do not consider such examples.}. Thus, it is always profitable to have at the end as little gas in the storage as possible within the constraints boundaries. It means that the optimal trajectory has to terminate at the lower boundary, and hence the problem can be considered as having the fixed terminal state. If we variate the forward curve $F\to F+\delta F$, this will lead to variation of the optimal strategy. However, the new strategy preserves the terminal level.

There are examples of the contracts for which this logic is not applicable. For instance, if the remaining gas can be sold at the fixed price $F_e$, then the derivative of the storage value with respect to the terminal state is given by Eq.~(\ref{eq:4.21})
\[ \frac{\p S}{\p q_{end}} = F_e - C = 0 \]
from which the condition (\ref{eq:4.22})
\[C = F_e\]
follows. The vanishing derivative implies that the trajectory may terminate anywhere between the upper and lower boundaries. The variation of the forward curve would lead to an ``arbitrary'' variation of the optimal trajectory which does not preserve the terminal state.

Another more common example is a swing option. Usually a swing option is a gas supply contract. The option holder is given a right to purchase gas from a producer at some ``contract'' price. He would then sell the gas on the market at the ``market'' price. Thus, the effective price for the option holder is the spread between the ``contract'' and the ``market'' price. This spread can be positive or negative. A typical swing contract obliges the option holder to take some amount of gas within the contract period, however the option holder has some flexibility: the total taken amount can be anything within the constraints -- minimal and maximal total taken volumes. The option holder also has some flexibility during the contract time to take bigger or smaller amount of gas on particular days, making maximum of the opportunities to earn on the spread. The swing contract can be easily formulated in terms of a storage, and hence possesses all features of a storage option. However the matter of fact that the spread between contract and market prices can be negative, makes the swing option having slightly different peculiarities.

The difference between the swing and storage option is rather relative. One can find examples of swing options having all the features of a storage option. If the swing option is deep in the money (i.e. the market prices are much higher than the contract prices), then the option holder is most likely to take the maximal allowed amount of gas, making the optimal trajectory to terminate at the lower boundary. Similarly, if the option is deeply out of the money, the optimal trajectory is very likely to terminate on the upper boundary. However if the swing option is at the money, it is quite probable that the total taken amount of gas will be between maximum and minimum. In this case the swing option problem becomes a problem with free terminal state.

This section is devoted to the problem with the free terminal condition. We will show that this problem has a different time value and derive the time value in a similar way it has been done for the problem with fixed terminal state.

The problem with free terminal state has an important feature. As it was shown in the Sec.~(\ref{sec:free_vs_fixed_terminal_condition}), the derivative of the option value with respect to the terminal volume must vanish. This leads to the condition on the trigger price $C= F_e$. Since there is no final unit price in the swing contract we conclude that for the problem with free terminal condition
\[C = 0\,,\]
from which obviously the condition
\[\delta C = 0 \]
follows. These conditions will simplify the derivation of the option value evolution equation. In particular, the variation of the target function becomes (compare with Eq.~(\ref{eq:deltaS}))
\begin{align}
    \label{eq:deltaS_swing}
    \delta S = & -\int r(T) \,\delta F(T)\,dT + \frac12 \int r^{(1)}\, L_{uu} \,du \,.
\end{align}
The second term gives a non-vanishing part of the average option value increment. If we compare this formula with Eq.~(\ref{eq:deltaS}) we see that $\delta S$ of the problem with fixed terminal state contains an additional negative term. Hence the averaged growth rate of the option value of the problem with free terminal state is higher than that of the problem with fixed terminal state, as expected.

The evolution equation for the option time value becomes
\[  dP(t) = \mu_p(t)\,dt + \sigma_p(t)\,dW_t\,;\]
with
\begin{align}
    \mu_p(t) = \frac{\la \delta S \ra_\dF }{\delta t}  = \frac{1}{2\,\delta t} \int r^{(1)}\, L_{uu} \,du\,.
\end{align}

Repeating the calculation in the previous sections we finally obtain
\begin{align}
    \label{eq:av_mu_swing}
    & \bar \mu_p(t) = \frac{\Delta r}{2} \sum_i \frac{\Lambda_i(t)}{|\dot F_{0i}(t)|}\,
\end{align}

\subsubsection{Example calculation}

Let the evolution of the spread be described by a Normal Brownian one-factor price process
\begin{align}
    \label{eq:brownian_motion}
    dF(t,T) = \kappa(t,T)\,dW_t\,;\qquad \kappa(t,T) =  \kappa_0\,e^{-\alpha (T-t)}\,;
\end{align}
Note that a natural choice for the price process describing the evolution of gas forward curves is a geometric Brownian motion, whereas for the spreads evolution a plane Brownian motion is a more suitable choice. Unlike the volatility $\sigma$ of the geometric Brownian motion, the Normal volatility $\kappa$ includes additional dimension of price.

\begin{figure}[htb!]
  \begin{center}
    \includegraphics[angle= 0, width = 0.8\textwidth] {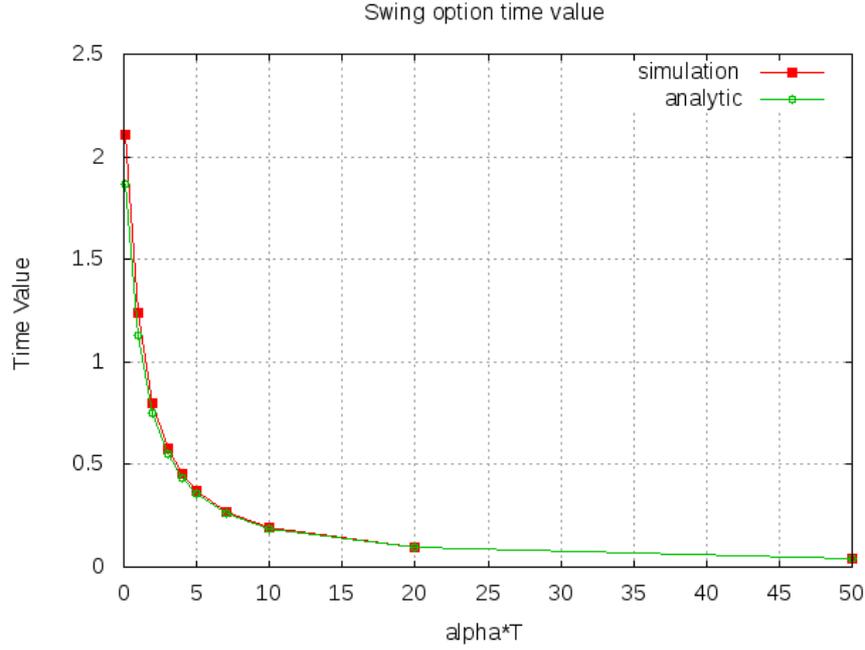}  
    \caption{Time Value $\Phi$ as a function of $\alpha\,T_e$. Analytic solution and numerical simulation.}
    \label{fig:4}
  \end{center}
\end{figure}

The correlation function of the price process~(\ref{eq:brownian_motion}) becomes
\begin{align}
    \Lambda_{ij}(t) = \frac{1}{\delta t}\,\la dF(t,T_i)\,dF(t,T_j)\ra  = \kappa_0^2\,e^{-\alpha(T_i-t)}\,e^{-\alpha(T_j-t)}\,.
\end{align}
Let the initial forward curve be given by
\begin{align}
    F_0(T) = \Delta F\,\sin(\omega \,T)\,;\qquad \omega = \frac{\pi\,N}{T_e} = \frac{\pi}{\Delta T}\,,
\end{align}
Repeating the calculation similar to that in the previous section we obtain the swing option time value
\begin{align}
    \label{eq:swing_option_value}
    V_T = \frac{\Delta r\,\kappa_0^2\,T_e^2}{8\,\pi\,\Delta F}\,\Phi(\alpha\,T_e)\,;\qquad\text{where} \quad
    \Phi(x) = \frac{e^{-2\,x}+2\,x-1}{x^2}\,.
\end{align}
This formula resembles the formula of the time value of a storage option~(\ref{eq:storage_time_value}), where the volatility $\kappa$ replaces the product $F_c\,\sigma_0$. The asymptotics of the swing option time value for the big mean reversion parameter coincides with that for the storage option:
\begin{align}
    V_T \approx  \frac{\Delta r\,\kappa_0^2}{4\,\pi\,\Delta F}\,\frac{T_e}{\alpha}\,,\qquad \alpha\,T_e\to\infty\,.
\end{align}
However in the limit of small mean reversion the swing option time value reveals different properties:
\begin{align}
    V_T \approx   \frac{\Delta r\,\kappa_0^2\,T_e^2}{8\,\pi\,\Delta F}\,\Big(2-\frac{4}{3}\alpha\,T_e\Big)\,,\qquad \alpha\,T_e\to 0\,.
\end{align}
For $\alpha=0$ the time value reaches its maximum
\begin{align}
    V_T^* = \frac{\Delta r\,\kappa_0^2\,T_e^2}{4\,\pi\,\Delta F}
\end{align}
and on the entire range the time value is a decreasing function of $\alpha$ (See Fig.~\ref{fig:4}).

This result can be easily interpreted. The main driver of the storage option time value is the change of exercise strategy. Since any change of the exercise strategy must preserve the terminal volume level, the storage option becomes a zero sum game: any decision to inject more at some time moment must be compensated by the decision to release more at some other time moment. Hence the storage option is in some sense similar to a set of time spread options. The only way the time spread option can increase its value is an asynchronous movement of different parts of the forward curve. In the limit $\alpha\to 0$ the forward curve moves in a parallel way (in log scale), and hence does not lead to any change of the exercise strategy, i.e. to the increase of the time value.

On the contrary, the swing option is similar to a strip of european call options. Indeed, since there is no terminal level preservation condition, any change of the strategy at one (delivery) time has no impact on the rest of the trajectory. If for instance we hedge the exercise trajectory at the very first observation time moment $t=0$, then the change of the exercise decision for any time $T$ would make a profit. The pay-off of the exercised volume on the time $t$ will have a structure of the european vanilla option. The effective square volatility of the option with exercise time $T$ is given by
\[
    \bar\varsigma^2 = \int_0^T \sigma^2(\tau,T)\,d\tau\,.
\]
It's easy to see that the biggest effective volatility $\bar\varsigma^2$ is achieved for the flat volatility term structure~$\alpha=0$, thus leading to the highest possible time value of the swing option.

This has a strong impact on the way the price process for the forward curve evolution should be calibrated. Let us consider a price process in the form
\begin{align}
    dF(t,T) = \kappa_1(t,T)\,dW_t^{(1)} + \kappa_2(t,T)\,dW_t^{(2)}\,;
\end{align}
where $dW_t^{(1)}$ and $dW_t^{(2)}$ are (possibly correlated) standard Brownian processes. Let also $\kappa_1$ and $\kappa_2$ be  exponential functions, describing the short term and long term volatilities
\[
    \kappa_1 = \kappa_{10}\,e^{-\alpha_1\,(T-t)}\,;\qquad \kappa_2 = \kappa_{20}\,e^{-\alpha_2\,(T-t)}\,.
\]
Here the mean reversion parameter $\alpha_1$ is relatively big (compared to $1/T_e$), i.e. the term $\kappa_1$ describes the short term volatility. The second term $\kappa_2$ describes the long term volatility, which means $\alpha_2\ll 1/T_e$. The correlation function decouples into three terms
\begin{align*}
    & \Lambda_{ij} = \frac{1}{dt}\,\la dF(T_i)\,dF(T_j)\ra =
    \Lambda_{ij}^{(1)} + \Lambda_{ij}^{(12)} + \Lambda_{ij}^{(2)}\,, \qquad \text{where}   \\
    & \Lambda_{ij}^{(1)} =  \la F(T_i)\,F(T_j)\ra\,\kappa_1(T_i)\,\kappa_1(T_j)\,; \\
    & \Lambda_{ij}^{(2)} =  \la F(T_i)\,F(T_j)\ra\,\kappa_2(T_i)\,\kappa_2(T_j)\,; \\
    & \Lambda_{ij}^{(12)} = \rho\,\la F(T_i)\,F(T_j)\ra\,
        ( \kappa_1(T_i)\,\kappa_2(T_j) + \kappa_2(T_i)\,\kappa_1(T_j))\,.
\end{align*}
Since the option time value (both storage and swing) is a linear functional on the correlation function, the time value also splits into three components, which can be estimated independently.

For evaluation of the storage option, the second and third terms will give no contribution, since they correspond to the long term volatility ($\alpha\to 0$). Consequently for the storage option the second term in the r.h.s. of the price process can be neglected.

This is not the case for the swing option. Since both -- long term and short term volatilities contribute to the option time value, we can not neglect the long term volatility in the price process.

\section{Discussion}
\label{sec:discussion}

In Sec.~\ref{sec:Deterministic_problem} we considered a deterministic storage problem and have fount its solution in an implicit form. The solution provided by Eq.~(\ref{eq:solution}) contains a free parameter -- trigger price, which can be found from boundary conditions. The obtained solution reveals the key features of the intrinsic optimal exercise strategy.
\begin{itemize}
    \item The optimal exercise is bang-bang. There is a ``dead zone'' around the trigger price. The optimal strategy is to inject at maximum rate if the price is below the zone, release at maximum rate if the price is above the dead zone and do nothing if the price is within the dead zone.
    \item The width of the dead zone is defined by the operating costs.
    \item The trigger price is constant for every piece of the trajectory where it does not sticks to the boundary. Different pieces of trajectory separated by the boundary touch may have different trigger level. If the trajectory touches the boundary in the interval $t\in(0,T_e)$, the conditions~(\ref{eq:bc1}-\ref{eq:bc4}) must be satisfied.
\end{itemize}

We also considered an impact of different constraints on the solution. In particular we found that
\begin{itemize}
    \item the cycle constraint is equivalent to additional injection/release costs.
    \item the carry cost preserves the bang-bang property, but makes the trigger price growing with the time at the rate of the carry cost.
    \item The volume dependent injection/release costs lead to a time-dependent trigger level, but preserve the bang-bang property.
    \item Fir the price of the underlying is strictly positive, and there is no final unit price, then the storage problem (both intrinsic and stochastic) can be considered with fixed terminal condition. If there is a positive final unit price, or if the prices of the underlying can be negative, then the optimisation problem becomes a problem with free terminal condition. This problem has an additional condition on the trigger level~(\ref{eq:4.22}) for storage or~(\ref{eq:4.23}) for swing option.
\end{itemize}

In Sec.~\ref{sec:stochastic_problem} we developed a perturbation analysis to find the solution of the stochastic problem in rolling intrinsic approximation.

First important result obtained in this section finds the relation between the intrinsic and stochastic solutions. In particular we show that the stochastic optimal exercise has exactly the same bang-bang feature with a trigger level and a dead zone. The stochastic trigger level is given by the averaged intrinsic trigger price (Eq.~(\ref{eq:stochastic_trigger_price_1})), where the average is taken over all realisations of the spot price process. Under the condition~(\ref{eq:condition_for_rolling intrinsic}) the average trigger price equals approximately the current intrinsic trigger price (conditional on the current initial condition and forward curve). This condition defines the range of applicability of the rolling intrinsic approximation.

Next we show that the hedge position has no impact on the expected option value. The only relevant part of the hedge strategy is the instant (prompt) exercise at the current time. This justifies the assumption that the intrinsic exercise profile as a good approximation of the optimal stochastic hedge profile, and generally, the rolling intrinsic strategy is a reasonable approximation for the optimal stochastic strategy.

By applying a variational analysis to the intrinsic solution we obtain the option greeks -- delta and gamma. This allows us to find the option time value. Making use of the option gamma we derive a stochastic differential equation governing the evolution of the option value. The drift term (Eq.~(\ref{eq:av_mu_storage}) for storage and Eq.~(\ref{eq:av_mu_swing}) for swing option) can be integrated yielding the option time value.

We apply the obtained time value formula for a toy examples of the storage and swing contracts with a simple 1-factor price process. The storage option time value (Eq.~(\ref{eq:storage_time_value})) is a bell-shaped function of $\alpha$, approaching zero for $\alpha\to 0$ and $\alpha\to\infty$, and reaching its maximum at $\alpha\,T_e\approx 5$. The swing time value appears to be a decreasing function of $\alpha$ for the entire range, reaching the maximum for $\alpha=0$ and approaching zero for $\alpha\to\infty$.

The option time value appears to be a linear functional on the prices correlation function. From this fact we conclude that if the correlation function can be decoupled into independent components, the option time value can also be decoupled. This allows us to investigate effect of different components of the price process on the time value independently.

The difference in the limiting case $\alpha\to 0$ between swing and storage option time value has an important consequence. This limiting case corresponds to the ``long term'' volatility due to the flat shape of the correlation function in this limit. If we consider a price process, whose correlation function could be decoupled into short and long terms, the impact of the long term component can be investigated independently. We conclude that the long term volatility has a significant impact on the swing option time value, but has no impact on the storage option time value. This fact can also be used in the price process calibration procedure.

Last important comment concerns the applicability of the obtained results. Throughout the derivation process we used a number of simplifying approximations. In particular, the rolling intrinsic approximation is valid for sufficiently small volatility. For the derivation of the option time value we also have neglected the surface effects (possible influence of the storage being completely full or empty), the bid-offer and operating costs, and used rather simple time-independent option constraints. For the derivation of analytic formulas of the time value we also implicitly used an assumption about sufficient smoothness of the forward curve, which justified the use of Taylor expansion. Another significant approximation was made for the definition of the swing option, where we have allowed the option holder to take any possible amount of the underlying limited only by the maximum release rate. 

All these approximations restrict the applicability of the obtained results. However all the qualitative results remain valid for much bigger range of parameters. And the main results can still be used for understanding the option value driving mechanisms and for building an intuition about the influence of different factors on the option value. 

Suppose that we are considering a real storage and we are interested in the applicability of the option value formula. For instance, we find that the intrinsic strategy spends a significant time on the boundary. This means that we can't neglect the surface effects, and the option value formula is not applicable directly. However, as the experiments show, all the major qualitative results still remain valid, and the option value formula could be corrected by some multiplier, which is weekly dependent on the storage parameters. Thus evaluating the storage for one set of parameters, we can easily predict how the value would change for a different set of parameters.

Another example of the restricted usability of the option value formula is the swing contract. A real swing contract usually has two global constraints -- maximum and minimum volume which can be taken during the contract period. These constraints were not taken into account in the swing value formula, which allowed us to use the free terminal condition approximation. However, if the option is ``at the money'', the probability is very high that the total taken volume will be between the maximum and minimum, and thus satisfy the requirement of the free terminal condition. Hence the swing formula remains a good approximation for the at-the-money option. The swing options which are either deep in- or out-of-the-money, are very likely to finish on either lower or upper boundary of the total taken volume. Such an option satisfies the condition with fixed terminal level, and hence a storage formula can be used to estimate the time value of such an option. An intermediate case, when the swing option is only slightly in- or out-of-the-money, neither storage nor swing formula become directly applicable. However, one can see that these two formulas allow a continuous transition, since they only differ by one term, which is not present in the swing formula. Thus the ultimate formula could be easily adjusted for the particular swing option.

\vspace{1cm}

{\LARGE \bf Appendix}

\appendix

\section{Derivation of Constraint Surface equation}
\label{ap:Contraint_surface}

To find the equation of the constraint surface we vary the equation \ref{eq:constraint}. Since $F(t,T)$ follows a Wiener type stochastic process, so does the $C(t)$. Hence we find the variation of the $Q_{end}$ up to the second order in $\delta F$ and $\delta C$:
\begin{align}
    dQ_{end} = & \Big(\dot q - r(t)\Big)\,dt +
    \frac{\p Q_{end}}{\p C}\, \delta C +
    \int \frac{\delta Q_{end}}{\delta F(T)}\,\delta F(T)\, dT +
    \delta C \int \frac{\delta^2 Q_{end}}{\p C\,\delta F(T)}\,\delta F(T)\,dT + \nonumber \\
    & \hspace{1cm} +
    \frac12\,\frac{\p^2 Q_{end}}{\p C^2}\, \delta C^2 +
    \frac{1}{2} \iint \frac{\delta^2 Q_{end}}{\delta F(u)\,\delta F(v)}\,
    \delta F(u)\,\delta F(v)\, du\,dv = 0\,.
\end{align}
On the optimal intrinsic trajectory $\dot q(t) = r(t,t)$, and the first term in the r.h.s. vanishes.

We introduce the notation:
\begin{align}
    & r^{(1)}(t,T)   = \Delta r\, \delta(C(t)-F(t,T)) \,; \\
    & r^{(2)}(t,T) = \Delta r\, \delta'(C(t)-F(t,T))\,; \\
    & K(t) = \int_t^{T_e} r^{(1)}(t,T)\,dT\,; \\
    & M(t) = \int_t^{T_e} r^{(2)}(t,T)\,dT\,,
\end{align}
where $\delta(x)$ is the Dirac delta-function, and $\delta'(x)$ -- its derivative. Making use of Eq.~(\ref{eq:solution_for_dot_q}) and the notation above, we find
\begin{align}
    & \frac{\p Q_{end}}{\p C}     =  K(t) \,; \\
    & \frac{\p^2 Q_{end}}{\p C^2} =  M(t) \,; \\
    & \frac{\delta Q_{end}}{\delta F(T)} =  -r^{(1)}(t,T) \,; \\
    & \frac{\delta^2 Q_{end}}{\p C\,\delta F(T)} =  -r^{(2)}(t,T) \,; \\
    & \frac{\delta^2 Q_{end}}{\delta F(t_1)\,\delta F(t_2)} = r^{(2)}(t,t_1)\,\delta(t_1-t_2)\,.
\end{align}
We thus obtain a quadratic equation on $\delta C$:
\begin{align}
    \label{eq:quadratic_eq_for_dC}
    \delta C = &\frac{1}{K}
        \left( \int_t^{T_e} r^{(1)} \,\delta F\,dT  -
                \frac12 \int_t^{T_e} r^{(2)}\,\delta F^2 \,dT +
                \delta C \int_t^{T_e} r^{(2)} \,\delta F\,dT -
                \frac{M}{2}\,\delta C^2
          \right)\,.
\end{align}
Here for simplicity we omitted arguments of the functions: $r=r(t,T)$,  $\delta F = \delta F(t,T)$, $K = K(t)$,  $M = M(t)$ and $\delta C = \delta C(t)$.

We search the solution in form of a series. As a zeroth order solution we take
\[ \delta C_0 = 0\,.\]
The first order solution becomes
\[ \delta C_1 = \frac{1}{K}
        \left(
                \int r^{(1)} \,\delta F\,dT  -
                \frac12 \int r^{(2)}\,\delta F^2 \,dT
          \right)
\]
Substituting $\delta C_1$ into the r.h.s. of Eq.~(\ref{eq:quadratic_eq_for_dC}), and leaving only 1st and 2nd order terms we finally obtain
\begin{align}
    &\delta C = \frac{1}{K}
        \left(  \int r^{(1)} \,\delta F\,dT  -
                \frac12 \int r^{(2)}\,\delta F^2 \,dT +
                \frac{1}{K}\iint r^{(2)}_u\,r^{(1)}_v\,L_{uv} -
                \frac{M}{2\,K^2}\iint r^{(1)}_u\,r^{(1)}_v\,L_{uv}
          \right)\,; \\
    & \delta C^2 =\frac{1}{K^2}\iint r^{(1)}_u\,r^{(1)}_v\,L_{uv}\,,
\end{align}
where we have used the notation
\[ L_{uv} = \delta F(u) \, \delta F(v)\,. \]

\section{Portfolio dynamics and time value}
\label{ap:A2}

In this section we show how the option time value can be obtained from the dynamics of the portfolio components in the rolling intrinsic strategy.

We start with summarising some basic facts from the option pricing theory. Let $F$ be a price of some underlying with delivery at time $T$, and let $f(t,F)$ be a price of an option expiring on $T$. Let the price be following a geometric Brownian motion (for simplicity we disregard the interest rate), which in Ito representation reads:
\[
    \frac{dF}{F} = \sigma(t)\,dW_t\,.
\]
Using Ito rule we can find a differential of the option price:
\[
    df = \frac{\p f}{\p t}\,dt + \frac{\p f}{\p F}\,dF + \frac12 \frac{\p^2 f}{\p F^2}\,dF^2 =
    \left(\frac{\p f}{\p t} + \frac12\, \sigma^2\,F^2 \,\frac{\p^2 f}{\p F^2}\right)dt + \frac{\p f}{\p F}\,dF\,.
\]
According to the non-arbitrage condition average of this expression must vanish: 
\[\la df \ra = 0\,.\]
Noticing that $\la dF \ra = 0$ we obtain
\begin{equation}
    \label{eq:theta_and_gamma}
    \frac{\p f}{\p t}\,dt + \frac12\, \frac{\p^2 f}{\p F^2}\,dF^2 = 0\,.
\end{equation} 
or substituting $dF^2$ explicitly
\begin{equation}
    \label{eq:BS}
    \frac{\p f}{\p t} + \frac12\, \sigma^2\,F^2 \,\frac{\p^2 f}{\p F^2} = 0\,.
\end{equation} 
This equation (up to the interest rate terms) is known as Black-Scholes equation. For us it is essential that the operator
\[
    \hat B = \frac{\p }{\p t} + \frac12\, \sigma^2\,F^2 \,\frac{\p^2 }{\p F^2} = \hat\theta +  \frac12\, \sigma^2\,F^2 \,\hat \Gamma\,,\qquad\text{where}\quad
    \hat\theta = \frac{\p }{\p t}\,;\quad \hat \Gamma = \frac{\p^2 }{\p F^2}\,;
\]
applied to any martingale process vanishes:
\[
    \hat B\,f = 0\,.
\]
Designating the operator of the full differential as
\begin{align}
     d = dt\,\hat B + dF\,\frac{\p}{\p F} = 
     dt\,\frac{\p }{\p t} + dt\,\frac12\, \sigma^2\,F^2 \,\frac{\p^2 }{\p F^2} + dF\,\frac{\p}{\p F}
\end{align}
we find
\begin{align}
    \label{eq:df}
    &df =  dt\,\hat B\,f + dF\,\frac{\p f}{\p F} = \frac{\p f}{\p F}\,dF\,;
\end{align}
Next we derive the expression for the differential of the option Delta $d\frac{\p f}{\p F}$. To do that, we notice first that the operators $d$ and $\p/\p F$ do not commute. Using the obvious commutation rule
\[
    \left[d\,, \frac{\p}{\p F}\right] = 
    d\, \frac{\p}{\p F} - \frac{\p}{\p F}\,d = - dt \, \sigma^2\,F\,\frac{\p^2 }{\p F^2}
\]
we find
\begin{align}
    \label{eq:d2f}
    & d\,\frac{\p f}{\p F} = \frac{\p }{\p F} \,df - \sigma^2\,F\,\frac{\p^2 f}{\p F^2} \, dt =
    \frac{\p^2 f }{\p F^2}\,dF - \sigma^2\,F\,\frac{\p^2 f}{\p F^2} \, dt \,;
\end{align}

\subsection{Dynamics of the portfolio components}

Let us consider a standard situation when an owner of an option hedges it with a combination of linear products. The full portfolio consists of three components: an option expiring on $T$, a hedge sub-portfolio, consisting of linear products, and a cash account. Let $\Pi(t)$ be the value of the full portfolio at the observation time $t$. Then we have
\[
    \Pi(t) = f(t) + H(t) + P(t)\,,
\]
where $f(t)$ is the option value, $H(t)$ is the value of the hedge portfolio, and $P(t)$ is the balance of the cash account.

The dynamics of the option value $f$ is given by Eq.~(\ref{eq:df})
\[
    df = \frac{\p f}{\p F}\,dF\,.
\]

Next we find the dynamic equation for other portfolio components. Let us consider a hedge approach, when the hedge portfolio consists of some amount of underlying $h$, i.e. hedge portfolio contains only linear products. The value of the hedge portfolio equals the current price of the hedge volume:
\[
    H = F\,h\,.
\]
In a classical scheme we may think of the option $f$ as a vanilla option on some kind of shares. In this case the hedge consists of some amount of shares which is purchased prior the expiry date. The same logic can be applied to a commodity market. For instance $f$ could imply an option to purchase some amount of gas. In this case hedge consists of forward contracts with delivery on the option expiry date. Although strictly speaking, the cash-flow associated with the delivery on the forward contract can take place on a later stage, we assign the cash-flow to the current (observation) time $t$, and hence the value of the hedge portfolio reflects the total price of underlying, and not only of the forward contracts.

The hedge portfolio is a product of two stochastic variables. Using Eq.~(\ref{eq:ItoChainRule}) from the main text, we find
\begin{align}
    dH = h\,dF + (F + dF)\,dh.
\end{align}
The latter equation can be formulated as \emph{retarded action principle}. This principle follows from a physical meaning of the dynamic hedge: the change of the hedge volume takes place after the change of the price is observed, and the additional hedge volume $dh$ has to be purchased at the new price $F+dF$. Notice however that the retarded action principle does not require any special differentiation rule.

The dynamics of the cash-flow component is easy to find. The cash-flow appears due to the change of the hedge volume:
\begin{align}
    dP = -(F+dF)\,dh\,.
\end{align}

The increment of the portfolio value is given by
\begin{align}
    d\Pi = df + dH + dP = \left(  \frac{\p f}{\p F} + h \right)\,dF\,.
\end{align}

First we notice that if we use the delta-hedge, i.e. if the hedge volume equals the option delta with the opposite sign
\[h = - \frac{\p f}{\p F}\,,\]
then, as expected, the variance of the portfolio equals zero:
\[
    d\Pi = df + dH + dP\equiv 0\,.
\]
Next important observation can be made about the relation between hedge portfolio and cash-flow. Averaging the increments $dH$ and $dP$ with respect to the stochastic price increment~$dF$ we obtain:
\begin{align}
    \label{eq:dH}
    \la dH \ra_{dF}  &=  - \la dP \ra_{dF}\,,
\end{align}
where $\la\cdot\ra_{dF}$ is the averaging over the price increment.

\subsection{Exact delta-hedge}

Let us consider the case when the hedge volume equals the exact (negative) option delta
\[
    h = -f' = -\frac{\p f}{\p F}\,;\qquad H = -F\,f'\,,
\]
Using Eq.~(\ref{eq:d2f}) we get for the hedge increment:
\[
    dH = -f'\,dF - F\,df' - dF\,df' = - (f' + F\,f'')\,dF\,.
\]
Note that on average the hedge portfolio value and the cash account do not change:
\[
    \la dH \ra = 0\,; \qquad   \la dP \ra = 0\,.
\]

\subsection{Intrinsic target function and time value}
\label{app:timevaluefromtargetfunction}

Here we show how the time value of a vanilla option can be calculated from the intrinsic target function. Let $\Delta_{int}$ be the intrinsic option delta, i.e. the delta calculated at the option maturity time
\[
    \Delta_{int}(F) = \frac{\p f(F,T)}{\p F}\,.
\]
Similarly we define the intrinsic gamma
\[
    \Gamma_{int}(F) = \frac{\p^2 f(F,T)}{\p F^2}\,.
\]
The intrinsic target function $S_{int}$ is defined such that at every time moment it is equal to the intrinsic option value
\[
    S_{int}(F) = f(F,T)\,.
\]
The variation of the target function is given by
\begin{align}
    \delta S_{int} = \Delta_{int}\,\delta F + \frac12\, \Gamma_{int}\,\delta F^2\,.
\end{align}
Note that the increment $\delta F$ is independent, and hence
\[
    \la \delta S_{int} \ra =\frac12\, \la \Gamma_{int}\,\delta F^2 \ra\,.
\]

Integrating $\delta S_{int}$ over the time and averaging over the realisations of the stochastic price process, we get
\begin{align}
    \int_0^T \la \delta S_{int} \ra = \la S_{int}(F(T))\ra - S_{int}(F(0))\,.
\end{align}
Next we observe that $S_{int}(F(0))$ is the intrinsic option value at time $t=0$, and the expectation $\la S_{int}(F(T))\ra$ is nothing else but the option value at time $t=0$, and thus the latter integral equals the option time value $V_T$:
\begin{align}
    V_T = \int_0^T \la \delta S_{int} \ra = \frac12\int_0^T \la \Gamma_{int}\,\delta F^2 \ra \,.
\end{align}

\subsection{Intrinsic delta-hedge of the storage option}

A storage option can be considered as a sequence of simple trades of buying and selling some amount of underlying. The logic of the derivation of a vanilla option time value can be applied to the storage option, if we consider it as a strip of (correlated) vanilla options with a pay-off $-\dot q(T)\,F(T)\,dT$, where $\dot q(T)\,dT$ is the volume purchased at time $T$ for the delivery period $dT$ at the spot price $F(T)$.

For every delivery time $T$ there is a forward price $F(t,T)$ which is as a function of the observation time $t$. Suppose we have an optimisation model, which calculates an intrinsic exercise profile $q_{int}(t,T)$ based on the current forward curve $F(t,T)$. Then $\dot q_{int}(t,T)$ is the volume of underlying (per delivery period $dT$) to be traded on time $T$. The value
\[
    s(t,T) = -\dot q_{int}(t,T)\,F(t,T)\,;\qquad t<T
\]
can be interpreted as a \emph{target function density} for the trade on the delivery time $T$. The same function after expiry is constant
\[
    s(t,T) = -\dot q_{int}(T,T)\,F(T,T)\,;\qquad t>T\,.
\]
Now the storage option target function can be expressed simply as an integral of the target function density over the delivery time
\begin{align}
    \label{eq:full_intrinsic_target_function}
    S(t) = \int_0^{T_e} s(t,T)\,dT = 
    -\int_0^t \dot q_{int}(T,T)\,F(T,T)\,dT - \int_t^{T_e} \dot q_{int}(t,T)\,F(t,T)\,dT
\end{align}
The first integral in this expression represents the value of the closed trades, whereas the second integral represents the future profit calculated with the current forward curve and exercise profile. At the last time moment $t=T_e$ the second integral vanishes, and the target function equals the cumulative cash flow from the closed trades. Note that the cumulative cash flow only depends on the exercise volumes $\dot q_{int}(T,T)$, and not on the hedge volumes $\dot q_{int}(t,T)$ for $t<T$.

Now we apply the logic of the section~\ref{app:timevaluefromtargetfunction} in order to calculate the storage option time value. Obviously the target function calculated at the time $t=0$ yields the intrinsic storage option value
\[
    S(0) = V_{int}\,.
\]
The expectation value of the terminal target function value gives the true option value
\[
    \la S(T_e) \ra = V
\]
(this statement relies on the fact that the intrinsic exercise is an optimal prompt exercise, which has been justified in the Sec.~\ref{sec:4.6.4}), and hence, the time value can be obtained as
\begin{align}
    V_T = \int_0^{T_e} \la dS \ra\,,
\end{align}
where $dS$ is a full differential of the target function. 

Some comment should be made regarding the calculation of the target function differential. First we split the target function into two parts:
\[
    S = S_c + S_o\,,
\]
where
\begin{align}
    S_c = -\int_0^t \dot q_{int}(T,T)\,F(T,T)\,dT
\end{align}
represents the ``closed'' trades, and
\begin{align}
    S_o = - \int_t^{T_e} \dot q_{int}(t,T)\,F(t,T)\,dT
\end{align}
represents the open trades, i.e. the remaining ``future'' option value. The definition of the target function in the main text coincides with the second ``open'' component of the target function defined above. However we can show that the open target function can also be used for the derivation of the time value.

The variation of the target function should be done with respect to the forward curve $F$ and time~$t$. The target function $S$ is an explicit function of the forward curve and of the time, and the dependence on the time is due to the limits of the integrals. It is easy to see that the time variance of the closed component equals in absolute value the time variance of the open component but has an opposite sign:
\[
    \frac{\p S_c}{\p t}  + \frac{\p S_o}{\p t} = 0\,.
\]
Thus the target function can be considered as an explicit function of the forward curve only. The full differential includes only the differentiation with respect to the forward price:
\[
    dS = \int \frac{\delta S}{\delta F}\,dF(u)\,du + 
    \frac12\, \int \frac{\delta^2 S}{\delta F(u)\,\delta F(v)}\,dF(u)\,dF(v)\,du\,dv
\]
The closed part of the target function is constant with respect to the forward curve variation (the variation of the forward curve impacts only the future values of the curve $F(t,T)$ for $T>t$), and hence
\begin{align}
    dS =  \eth S_o\,,
\end{align}
where we designated $\eth$ the variation which should be calculated for constant time $t$. Thus
\begin{align}
    V_T = \int_0^{T_e} \la \eth S_o \ra\,.
\end{align}

It is interesting to note that for the storage option the hedge value coincides with the (negative) open target function:
\[
    H(t) = \int_t^{T_e} \dot q_{int}(t,T)\,F(t,T)\,dT = -S_o\,,
\]
and hence
\begin{align}
    \la \eth S_o \ra = -\la \eth H \ra = \la dP \ra
\end{align}
where $P$ is the cumulative cash flow from the hedge trades. Note that this relation would not be valid for a vanilla option, since in general case the target function can not be expressed as $q_{int}\,F$.

\subsubsection{Cumulative cash flow}

There is another way to see how the time value can be obtained from the cumulative cash flow. By definition the cash flow (resulting from the hedge trades) with delivery on $T$ per delivery interval $dT$ is given by
\[
    dP_T(t,T) = - (F + dF)\,dh\,; \quad t<T\,; \qquad \text{where}\quad h = \dot q_{int}\,.
\]
\[
    dP_T(t,T) = 0\,,\quad\text{for}\quad t>T\,.
\]
The value $P_T$ has a meaning of the cash flow density over the delivery time. The total cash flow increment at time $t$ is related to $dP_T$ as
\[
    dP(t) = \int_0^{T_e} dP_T(t,T)\,dT\,,
\]
from which the total cash flow follows as
\begin{align}
    P(t) = \int_0^t d\tau \int_\tau^{T_e} dP_T(\tau,T)\,dT = \int_0^t d\tau \int_0^{T_e} dP_T(\tau,T)\,dT\,.
\end{align}
Both $P_T$ and $P$ should be considered as functions of time $t$ with $T$ been a parameter.

Integrating $dP_T(t,T)$ from 0 to $T$ and noticing that $P_T(0) = 0$, we get the cumulative cash flow per delivery period $[T, T+dT]$:
\[
    P_T(T) = \int_0^T dP_T = - \int_0^T (F + dF)\,dh\,.
\]
Now we use the integration by parts, which in our case can be represented as
\begin{align}
    \left. (F\,h)\right|_0^T = \int_0^T h\,dF + \int_0^T F\,dh + \int_0^T dF\,dh
\end{align}
For the averaged cumulative cash flow we obtain
\[
    \la P_T(T) \ra = F(0)\,h(0) - \la F(T)\,h(T) \ra = 
     F(0)\,\dot q_{int}(0) - \la F(T)\,\dot q_{int}(T) \ra\,.
\]
The total average cash flow is given by integrating $\la P_T\ra$ over the delivery time:
\[
   \la P(T_e) \ra = \int_0^{T_e}\la P_T(T)\ra\,dT = \int \dot q(0,T)\,F(0,T)\,dT - \la \int \dot q(T,T)\,F(T,T)\,dT \ra
\]
Noticing that 
\[
   - \int \dot q(0,T)\,F(0,T)\,dT
\]
equals the option intrinsic value, and 
\[
    -\la \int \dot q(T,T)\,F(T,T)\,dT \ra
\]
equals the true option value, we conclude that
\begin{align}
    V_T = \la P(T_e) \ra = \int_0^{T_e} \la dP \ra\,.
\end{align}

\section{Averaging an integral with delta function}
\label{app:averaged_integral}

In this section we consider an averaging of the integrals of the type
\begin{equation}
    I(t) = \int_t^{T_e} \delta(C - F)\,\phi(T,F)\,dT\,,
\end{equation} 
where $F = F(t,T)$ is the forward curve -- a stochastic function of observation and delivery times, and $\phi$ is an arbitrary function of $T$ and $F$. We will be interested in the average of the integral over the stochastic prices $\la I(t)\ra_F$.

Due to the delta-function under the integral we can write
\begin{equation}
    I(t) = \int_t^{T_e} \delta(C - F)\,\phi(T,C)\,dT\,,
\end{equation}

Introducing the set of trigger times $\{T_i: F(t,T_i) = C\}$ we can calculate the integral $I(t)$ for every observation time:
\begin{equation}
    \label{eq:summation3}
    I(t) = \sum_{i>t} \frac{\phi_i}{|\dot F_i|}\,;\quad\text{where}\quad
    \phi_i = \phi(T_i,C)\,; 
    \ F_i = C\,;
    \ \dot F_i =\left. \frac{\p F(t,T)}{\p T}\right|_{T=T_i} \,;
\end{equation}
This integral is a stochastic variable. Indeed, the process $F(t,T)$ is stochastic, and hence, the set of trigger times as well as the derivatives $\dot F_i$ are stochastic.

The average over the stochastic function $F$ is given by
\begin{equation}
    \label{eq:averaged_delta_integral_1}
    \la I(t) \ra = \int_t^{T_e} \la \delta(C - F)\ra\,\phi(T,C) \,dT
\end{equation} 
Here we have assumed that $C$ is constant.

The average of the delta function can be easily found using Fourier transformation:
\begin{equation}
    \label{eq:delta_fourier}
    \delta(C-F) = \frac{1}{2\,\pi}\int e^{i\,\omega\,(C-F)}\,d\omega
\end{equation}
Making use of the cumulant expansion and keeping only the two first cumulants (which, strictly speaking, is only exact for Gaussian variables) we find the average of the exponential:
\begin{equation}
    \la e^{i\,\omega\,(C-F)} \ra \approx \exp\left(i\,\omega\,(C-F_0) - \frac12\,\omega^2\,\sigma^2 \right)\,,
\end{equation} 
where we used
\[
    \la F \ra = F_0\,;\quad \lla F^2 \rra = \sigma^2\,.
\]
Substituting the averaged exponential into Eq.~(\ref{eq:delta_fourier}) and performing integration we obtain
\begin{equation}
    \la \delta(C-F) \ra = \frac{1}{\sigma\,\sqrt{2\,\pi}}\, \exp\left(
        - \frac{(C-F_0)^2}{2\,\sigma^2}
    \right)\,.
\end{equation} 

For the averaged integral we get:
\begin{equation}
    \label{eq:averaged_delta_integral_2}
    \la I(t) \ra \approx \int_t^{T_e} \frac{1}{\sigma\,\sqrt{2\,\pi}}\, \exp\left(
        - \frac{(C-F_0)^2}{2\,\sigma^2}
    \right)\, \phi(T,C) \,dT\,.
\end{equation} 
Note that both $F_0$ and $\sigma$ are functions of $t$ and $T$: 
\[ F_0 = F_0(t,T)\,;\quad \sigma = \sigma(t,T) \]
The exponent under the integral reaches its maximum at the trigger times $\{T_i\}$, when $F_0(T_i) = C$. Expanding the forward curve around this value
\[F_0(T) \approx C + \dot F_0(T_i)\,(T - T_i)\]
we notice that for each trigger time averaged delta function becomes a Gaussian bell (as a function of time $T$) of the width $\sigma_t$, which can be found from an approximate relation
\[
    \sigma_t(T_i) \approx \frac{\sigma(T_i)}{|\dot F_0(T_i)|}
\]

\subsection{slow $\phi$ approximation}

Suppose that the trigger times are distributed rather uniformly on the time axis. Then we may speak about characteristic time distance between the trigger times. Let $\Delta T$ be the average trigger time step
\[
    \Delta T \sim  \overline{ ( T_{i+1} - T_i) }\,.
\]
We say that function $\phi(T,C)$ is slow if it changes insignificantly within the average trigger time step:
\[
    \frac{\dot \phi}{\phi} \ll \frac{1}{\Delta T}\,,\qquad \text{where}\quad
    \dot \phi = \frac{\p}{\p T}\phi(T,C)\,.
\]
Let us also suppose that the time derivative $\dot F_i$ can be considered as constant plus maybe a small random correction:
\[
    |\dot F_i| = \dot F_0 + \epsilon_i\,,\qquad |\epsilon_i| \ll \dot F_0\,.
\]
In this case we can replace the summation in Eq.~(\ref{eq:summation3}) with the integral:
\begin{equation}
    I(t) = \int_t^{T_e} \delta(C - F)\,\phi(T,F)\,dT = \sum_{i>t} \frac{\phi_i}{|\dot F_i|}\approx
    \frac{1}{|\dot F_0|} \int_t^{T_e} \phi(T,C)\,dT
\end{equation} 
where we have assumed that the derivatives $|\dot F_0|$ can be considered as constant.

The same approximation can be used for the averaged integral (\ref{eq:averaged_delta_integral_2}). The biggest contribution to the integral is made by the intervals of the width $\sim 2\,\sigma_t$ around the trigger times
\[
    \la I\ra \approx \sum_i \int_{T_i-\sigma_t}^{T_i+\sigma_t} \frac{1}{\sigma\,\sqrt{2\,\pi}}\, \exp\left(
        - \frac{(C-F_0)^2}{2\,\sigma^2}
    \right)\, \phi(T,C) \,dT\,.
\]

If the volatility of the forward price process is not too big, so that
\[\sigma_t \lesssim  \Delta T \]
then the function $\phi$ is slow with respect to the width of the Gaussian bell around the trigger times in the integral~(\ref{eq:averaged_delta_integral_2})
\[
    \frac{\dot \phi}{\phi} \ll \frac{1}{\sigma_t}
\]
In this case the function $\phi$ can be roughly considered as constant on every integration interval $(T_i - \sigma_t, T_i+\sigma_t)$. 
We thus obtain
\[
    \la I(t)\ra \approx \sum_i \phi(T_i,C) \int_{T_i-\sigma_t}^{T_i+\sigma_t} \frac{1}{\sigma\,\sqrt{2\,\pi}}\, \exp\left(
        - \frac{(C-F_0)^2}{2\,\sigma^2}
    \right) \,dT\,.
\]
The volatility $\sigma$ as a slow function of the delivery time and can be considered as constant under the integral. Substituting in the first order around the trigger price $F_0(T)\approx C + \dot F_0 \,(T-T_i)$ we can estimate the integral over one trigger time as $1/|\dot F_0(T_i)|$, and the averaged integral becomes
\begin{equation}
    \la I(t)\ra \approx \sum_i \frac{\phi(T_i,C)}{|\dot F_0(T_i)|}
\end{equation} 
This means that even for the averaged integral under the specified assumptions the same summation formula can be used as for the stochastic integral~(\ref{eq:summation3}).

If we demand that the function $\phi(T,C)$ is a slow function of time not only with respect to the characteristic time $\sigma_t$ but also with respect to the time interval $\Delta T$
\[ \left|\frac{\phi(T_{i+1},C) - \phi(T_i,C)}{\phi(T_i,C)}\right| \ll 1 \]
and if $|\dot F_0(T_i)|$ can be considered as constant, then we can roughly estimate
\begin{align}
    \label{eq:integral_approximation}
    \la I(t)\ra \approx \frac{1}{|\dot F_0|} \int \phi(T,C)\,dT\,.
\end{align}
This approximation is used in the example calculation of the storage option time value.

\section{Some formulas used in the main text}
\label{ap:A1}

For the arbitrary sufficiently smooth functions $f(t)$ and $\varphi(t)$ the following identities are easy to prove:
\begin{align}
    \label{eq:A1}
    &\int_{-\infty}^{\infty} \delta(\varphi(t))\,f(t)\,dt =
            \sum \, \frac{f(t_i)}{|\dot \varphi(t_i)| } \,;  \\
    \label{eq:A2}
    & \int_{-\infty}^{\infty} \delta'(\varphi(t))\,f(t)\,dt =
            -\int \delta(\varphi(t))\, \frac{d}{dt} \left[ \frac{f(t)}{\dot \varphi(t)} \right] dt   =
            - \sum \, \frac{1}{|\dot \varphi(t_i)|}\, \frac{d}{dt}
                \left[ \frac{f(t)}{\dot \varphi(t)} \right]_{t=t_i} \,;  \\
    & \int_{-\infty}^{\infty} \delta''(\varphi(t))\,f(t)\,dt =
            \sum \, \frac{1}{|\dot \varphi(t_i)|}\,\frac{d}{dt}\, \frac{1}{\dot \varphi(t_i) }\,\frac{d}{dt}
                \left[ \frac{f(t)}{\dot \varphi(t)} \right]_{t=t_i} \,;
\end{align}
where the sum is over all points $\{t_i: \varphi(t_i)=0\}$.

In particular for the case $\varphi(t) = C-F(t)$ we have
\begin{align}
    \label{eq:A4}
    &\int_{-\infty}^{\infty} \delta(C-F(t))\,f(t)\,dt =
            \sum \, \frac{f(t_i)}{|\dot F(t_i)| } \,;  \\
    \label{eq:A5}
    & \int_{-\infty}^{\infty} \delta'(C-F(t))\,f(t)\,dt =
            \sum \, \frac{1}{|\dot F(t_i)| }\, \frac{d}{dt}
                \left[ \frac{f(t)}{\dot F(t)} \right]_{t=t_i} \,;
\end{align}


Making use the definitions from the main text (we omit the index $t$ where it is unambiguous)
\begin{align}
    &r(t,T)   = \Delta r\,\delta(C(t) - F(t,T)) = \Delta r\,\delta(C - F(T))\,; \\
    &r^{(1)}(T)       = \Delta r\,\delta'(C - F(T))\,; \\
    &r^{(2)}(T) = \Delta r\,\delta''(C - F(T))\,; \\
    & K = \int_t^{T_e} r^{(1)}(T)\,dT = \Delta r\, \sum \frac{1}{|\dot F(T_i)| }\,; \\
    & M = \int_t^{T_e} r^{(2)}(T)\,dT = -\Delta r\, \sum \frac{\ddot F(T_i)}{\dot F^2(T_i) }\,;
\end{align}
it is easy to prove the following identities:
\begin{align}
    & \int r^{(1)}(u)\,F(u)\,du = C\,K\,; \\
    & \int r^{(2)}(u)\,F(u)\,du = K + C\,M\,; \\
    & \int r^{(2)}(u)\,F(u)\,f(u) \,du = \int r^{(1)}(u)\, f(u)\,du + C\int r^{(2)}(u)\,f(u)\,du\,; \\
    & \int r^{(2)}(u)\,F(u)\,L(u,u)\,du = \int r^{(1)}(u)\, L(u,u)\,du + C\int r^{(2)}(u)\,L(u,u)\,du\,; \\
    & \iint r^{(2)}(u)\,r^{(1)}(v)\,F(u)\,L(u,v)\,\,du\,dv = \iint r^{(1)}(u)\,r^{(1)}(v)\, L(u,v)\,\,du\,dv + \nonumber \\
    & \hspace{6cm} + C\iint r^{(2)}(u)\,r^{(1)}(v)\,L(u,v)\,\,du\,dv\,,
\end{align}
where $f(u)$ is an arbitrary sufficiently smooth function.

\end{document}